\documentclass[sn-nature]{sn-jnl}

\usepackage{fullpage}
\usepackage{graphicx}
\usepackage{multirow}
\usepackage{amsmath,amssymb,amsfonts}
\usepackage{amsthm}
\usepackage{mathrsfs}
\usepackage[title]{appendix}
\usepackage{xcolor}
\usepackage{textcomp}
\usepackage{manyfoot}
\usepackage{booktabs,siunitx}
\usepackage{algorithm}
\usepackage{algorithmicx}
\usepackage{algpseudocode}
\usepackage{listings}
\usepackage{todonotes}
\usepackage{wrapfig}
\usepackage{hyperref}
\usepackage[utf8]{inputenc}
\usepackage[T1]{fontenc}
\usepackage{url}
\usepackage{nicefrac}
\usepackage{microtype}
\usepackage{tikz}
\usepackage{pgfplots}
\usepackage{subcaption}
\usepackage{enumitem}
\usepackage{pifont}
\usepackage{diagbox}
\usepackage{bm}
\usepackage{bbm}
\usepackage{mathtools}
\usepackage[nameinlink]{cleveref}
\usepackage{dutchcal} 

\theoremstyle{thmstyleone}%
\theoremstyle{thmstyletwo}%

\theoremstyle{thmstylethree}%

\hypersetup{
	colorlinks=true,
	linkcolor=myblue,
	citecolor=myblue,
	urlcolor=mygray,
}

\definecolor{myblue}{RGB}{0,0, 150}
\definecolor{mygray}{RGB}{90,90,110}

\def\method{%
  \textsc{Pano}%
  \futurelet\next\checknext
}

\def\checknext{%
  \ifx\next,%
  \else\ifx\next.%
  \else\ifx\next'%
  \else\ifx\next)%
  \else
    \ 
  \fi\fi\fi\fi
}
\begin{document}

\title[Article Title]{Physics-Aware Neural Operators for Direct Inversion in 3D Photoacoustic Tomography}

\author[1]{\fnm{Jiayun} \sur{Wang}}\email{peterw@caltech.edu}

\author[2]{\fnm{Yousuf} \sur{Aborahama}}\email{y.aborahama@caltech.edu}

\author[1]{\fnm{Arya} \sur{Khokhar}}\email{arya@caltech.edu}

\author[2]{\fnm{Yang} \sur{Zhang}}\email{zoengy@caltech.edu}

\author[1]{\fnm{Chuwei} \sur{Wang}}\email{chuweiw@caltech.edu}

\author[2,3]{\fnm{Karteekeya} \sur{Sastry}}\email{sdharave@caltech.edu}

\author[1]{\fnm{Julius} \sur{Berner}}\email{jberner@caltech.edu}

\author[2]{\fnm{Yilin} \sur{Luo}}\email{yilinluo@caltech.edu}

\author[4]{\fnm{Boris} \sur{Bonev}}\email{bbonev@nvidia.com}

\author[1]{\fnm{Zongyi} \sur{Li}}\email{zongyili@caltech.edu}

\author[4]{\fnm{Kamyar} \sur{Azizzadenesheli}}\email{kamyara@nvidia.com}

\author[2,3]{\fnm{Lihong V.} \sur{Wang}}\email{lvw@caltech.edu}

\author[1]{\fnm{Anima} \sur{Anandkumar}}\email{anima@caltech.edu}
\equaladv{Corresponding author.}

\affil[1]{\orgdiv{Department of Computing and Mathematical Sciences}, \orgname{California Institute of Technology}, \orgaddress{\street{1200 E California Blvd}, \city{Pasadena}, \postcode{91125}, \state{CA}, \country{United States}}}

\affil[2]{\orgdiv{Andrew and Peggy Cherng Department of Medical Engineering}, \orgname{California Institute of Technology}, \orgaddress{\street{1200 E California Blvd}, \city{Pasadena}, \postcode{91125}, \state{CA}, \country{United States}}}

\affil[3]{\orgdiv{Department of Electrical Engineering},  \orgname{California Institute of Technology}, \orgaddress{\street{1200 E California Blvd}, \city{Pasadena}, \postcode{91125}, \state{CA}, \country{United States}}}

\affil[4]{ \orgname{NVIDIA}, \orgaddress{\street{2788 San Tomas Express Way}, \city{Santa Clara}, \postcode{95051}, \state{CA}, \country{United States}} }


\abstract{
Learning physics-constrained inverse operators---rather than post-processing physics-based reconstructions---is a broadly applicable strategy for problems with expensive forward models. We demonstrate this principle in three-dimensional photoacoustic computed tomography (3D PACT), where current systems demand dense transducer arrays and prolonged scans, restricting clinical translation. We introduce \method (PACT imaging neural operator), an end-to-end physics-aware neural operator---a deep learning architecture that generalizes across input sampling densities without retraining---that directly learns the inverse mapping from raw sensor measurements to a 3D volumetric image. Unlike two-step methods that reconstruct then denoise, \method performs direct inversion in a single pass, jointly embedding physics and data priors. It employs spherical discrete-continuous convolutions to respect hemispherical sensor geometry and Helmholtz equation constraints to ensure physical consistency. \method reconstructs high-quality images from both simulated and real data across diverse sparse acquisition settings, achieves real-time inference and outperforms the widely-used UBP algorithm by approximately 33 percentage points in cosine similarity on simulated data and 14 percentage points on real phantom data. These results establish a pathway toward more accessible 3D PACT systems for preclinical research, and motivate future in-vivo validation for clinical translation.
}




\maketitle
\newpage

\section{Introduction}
Many computational imaging systems are governed by inverse problems of the form $\Psi = A P$, where $\Psi$ refers to measurements, $A$ is a large physics operator and $P$ is the latent image or field to be reconstructed. In practice, both reconstruction quality and runtime are limited by repeatedly applying and inverting $A$, especially when sensing is sparse, limited-view, or geometrically irregular. This bottleneck is central to wave-based imaging and is particularly acute in three-dimensional photoacoustic computed tomography (PACT). 
Photoacoustic computed tomography (PACT) has emerged as a powerful hybrid imaging modality that combines the optical contrast of diffuse optical tomography with the spatial resolution of ultrasonography \cite{wang2012photoacoustic,wang2016practical,wong2017fast}. By converting light absorption into ultrasonic waves through transient thermoelastic expansion, PACT serves as a noninvasive, high-resolution imaging modality at depths beyond the optical diffusion limit \cite{xia2014photoacoustic,zhou2016tutorial}. This unique capability enables detailed structural, functional and molecular imaging with rich intrinsic contrast and minimal speckle artifacts, making it complementary to other mainstream imaging modalities such as MRI, CT and X-ray imaging \cite{feinberg2012rapid,rahbar2013clinical}. More recently, several PACT systems with three-dimensional (3D) field-of-view (FOV) have been proposed \cite{lin2021high,cao2023single,brecht2009whole}, which outperform the 2D PACT system in terms of imaging depth and quality. 3D PACT enables various preclinical studies \cite{brecht2009whole,jathoul2015deep,gottschalk2019rapid} and clinical practice \cite{matsumoto2018visualising,oraevsky2018full}, with applications including 3D transcranial imaging \cite{huang2025fast} and whole body 3D imaging of live animals \cite{choi2023DLPACT}.

\begin{figure}[t!]
\vspace{-1em}
    \centering
    \includegraphics[width=0.9\columnwidth]{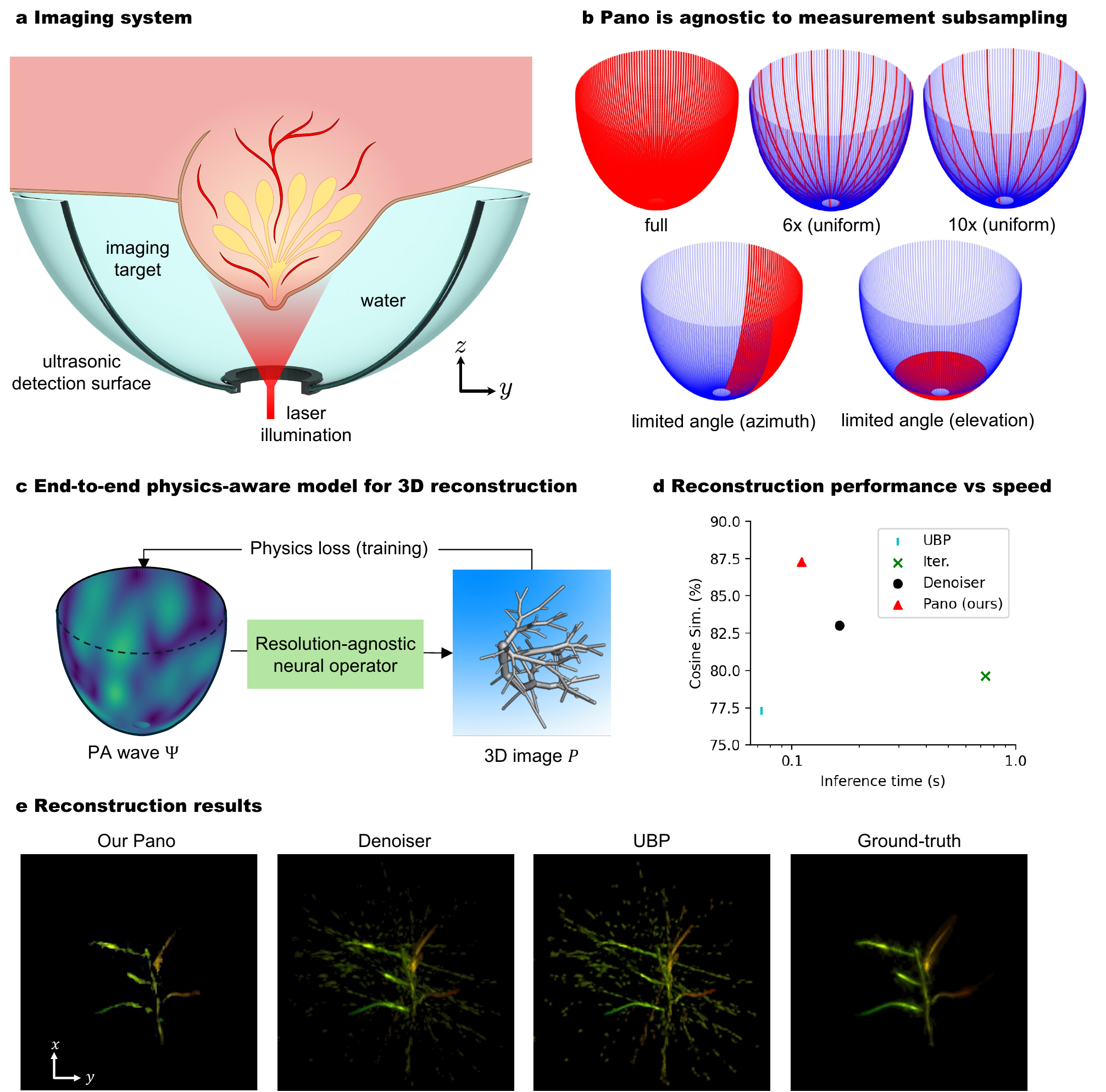} 
    \caption{  {\bf Overview: The proposed \textsc{Pano} (Photoacoustic imaging neural operator) reconstructs a 3D image (voxel) from photoacoustic radio-frequency (RF) data measurements}. 
    {\bf a,} Schematic diagram of the imaging system, which uses a hemispherical ultrasound (US) transducer array. The target is placed on top of the US detection surface of the transducer array and a laser illuminates the target. The photoacoustic (PA) waves are detected by the sensor for further processing and reconstruction with a data acquisition system.  
    {\bf b,} The arrangements of the transducer elements with subsampled measurement patterns (to accelerate the imaging): Full, uniformly subsampled measurements at 6$\times$, 10$\times$ in azimuth (top), and limited angle (3$\times$ acceleration) in azimuth and elevation. 
    {\bf c,} Overall architecture of the proposed deep learning framework \textsc{Pano}  for end-to-end 3D reconstruction. A neural operator is used to transform the PA wave $\Psi$ to 3D volumetric image $P$. A neural operator is designed to be agnostic to the sampling rate of the PA wave. As a cycle consistency check, reconstruction is further projected back to PA waves, and a physics loss is used to penalize if the reconstruction's PA wave deviates significantly from the input $\Psi$. 
     {\bf d,}  Reconstruction performance (cosine similarity) and inference speed of different reconstruction algorithms on real experimental data. The proposed neural operator \textsc{Pano} achieves improved reconstruction performance with faster inference time compared to baseline methods, such as deep-learning-based Denoiser \cite{choi2023DLPACT} and the iterative solver \cite{xu2002exact}. Compared with the well-adopted state-of-the-art reconstruction method UBP (universal back-projection algorithm) \cite{xu2005universal}, \textsc{Pano} achieves a 10 percentage points reconstruction performance gain with similar inference time. (Inference setting: $6\times$ uniform subsampling.)
     {\bf e,}  PA MAP image comparisons of different methods. The proposed \textsc{Pano} outperforms other methods, reconstructing 3D structures with higher fidelity and lower noise. MAP, maximum amplitude projection.
     }
    \label{fig:teaser}
\end{figure}

While advances in three-dimensional (3D) PACT systems have enabled impressive imaging performance, 
significant challenges remain. For example, a high-resolution 3D PACT system \cite{cao2023single,choi2023DLPACT} requires extensive resources, including dense transducer arrays 
and prolonged imaging times (e.g., a single 10-second breath-hold for breast imaging). These requirements impose limitations on imaging speed, cost, and patient comfort, especially in resource-constrained or clinical settings where reducing scan time and transducer count is paramount.

There is a growing interest in developing compressed sensing methods to accelerate PACT systems 
\cite{choi2023DLPACT}. Such methods aim to reconstruct high-fidelity and high-quality images with subsampled sensory data below the Nyquist limit. 
Classical compressed sensing~\cite{arridge2016accelerated,davoudi2019deep,zheng2023deep} accelerates PACT by assuming a sparse prior 
\cite{farnia2020dictionary}, relying on dictionary learning and using wavelet decomposition \cite{tzoumas2014spatiospectral}. The universal back-projection algorithm (UBP) \cite{xu2005universal} is one of the most common reconstruction algorithms in preclinical and clinical settings due to its balanced speed and performance. 
Recently, deep neural networks have shown great success in image denoising \cite{ronneberger2015u,rombach2021highresolution} and researchers are stacking the denoising network after the conventional solvers to improve the PACT image reconstruction quality further. 
While most work aims at removing artifacts in 2D PACT images \cite{davoudi2019deep,shahid2021deep}, some researchers propose deep learning methods for 3D PACT systems. Specifically, \cite{zhang2021deep} proposed an algorithm that converts the 3D problem into 2D by simulating and training data in the axial-elevation plane. \cite{choi2023DLPACT,zheng2023deep} introduced 3D fully-dense U-net to remove artifacts in  the 3D images. 
However, the aforementioned work operates on the reconstructed (3D) volumetric image and 
relies on another physics-based solver to reconstruct the image from the sensory data input. 
They act like a denoiser and have two disadvantages: 1) The reconstruction performance can be low as it is dependent on the physics-based solver, which provides the input to the denoiser; 2) The reconstruction time of the method can be long as the run time of the physics-based solver needs to be considered as well. 

{\bf Our approach:} In this work, we present a first end-to-end physics-aware neural operator framework for 3D PACT image reconstruction, \method (PACT imaging neural operator). 
 Unlike existing denoising networks \cite{zhang2021deep,choi2023DLPACT} that improve the solver reconstruction with a learned data prior, \method learns a direct inverse neural operator that better approximates the inverse of the forward acoustic operator, jointly embedding physics and data priors. This formulation improves generalizability: \method shows strong sim-to-real transfer while being primarily trained on simulation data.
Additionally, \method is a neural operator~\cite{kovachki2023neural}, which is agnostic to subsampling density within the same sensor array geometry and can adapt to different sparse acquisition patterns without retraining. 
\method thus substantially reduces the reliance on dense transducer arrays and prolonged scan time, with improved image reconstruction performance over existing methods. 
Our framework achieves high-fidelity image reconstruction using fewer transducers and limited scan angles, offering a cost-effective and fast alternative to traditional approaches, with potential as a foundation for future clinical translation.

The proposed model integrates data-driven learning with physics constraints to achieve robust and accurate reconstructions, even with noisy or incomplete data. Unlike conventional image-denoising methods that often decouple data priors from physics, our method preserves geometric relationships by leveraging the hemispherical transducer arrangement and learning directly on the hemispherical domain with spherical convolutions. 
Additionally, we introduce a sampling-based strategy to balance computational efficiency and gradient stability, enabling scalable implementation for large-scale data without sacrificing fidelity.

Key contributions of our approach include: (i) improved 3D reconstruction performance over the widely-adapted UBP solver~\cite{xu2005universal} by approximately 33 percentage points on simulated data and 14 percentage points on real phantom data, and over an existing deep learning method~\cite{choi2023DLPACT} by approximately 6 percentage points on simulated data and 11 percentage points on real phantom data; (ii) the ability to reconstruct high-quality images with only 33\% scan angle coverage; (iii) generalizability across simulated and real-world data through domain adaptation; and (iv) validation that the learned neural operator is resolution-agnostic~\cite{kovachki2023neural}: a single trained model maintains consistent performance across different sparse acquisition patterns of the same sensor array without retraining.
Our experiments demonstrate the feasibility of achieving high-resolution, real-time 3D imaging with significantly reduced system complexity and cost. This advancement enhances imaging speed and reduces hardware cost, establishing a foundation for future in-vivo studies that could broaden the accessibility of 3D PACT in preclinical and, ultimately, clinical settings.

\section{Results}
 A schematic of the 3D PACT imaging system used in the paper is depicted in Fig.~\ref{fig:teaser}a, which illustrates the source of illumination, the object being imaged (i.e., an adult human breast), the ultrasound coupling medium (water), and a hemispherical ultrasonic detection surface. 
 
To benchmark the accuracy of our approach we decided to use a hemispherical ultrasonic detection surface similar to the one in~\cite{lin2021high} (see Methods). 
We form initial-pressure maps as $p_0(\mathbf r)=\Gamma\,\mu_a(\mathbf r)\,\Phi(\mathbf r)$, with a simple homogeneous fluence $\Phi(\mathbf r)=\Phi_0$. Acoustic propagation is modeled in a homogeneous, lossless medium (no attenuation), and the forward operator is evaluated semi-analytically in the frequency domain under the free-space Green’s function for a homogeneous background. Time-domain detector signals are obtained by inverse FFT of the frequency-domain fields at the (point) detector locations on four replicated quarter-arc arrays matching the system geometry. We match the receive chain by band-limiting to the array response and the DAQ (7.5\,MHz analog anti-alias; 20\,MHz sampling), in addition to accounting for the transducers' sensitivity, then add additive white Gaussian noise (AWGN). We use the ground-truth $p_0$ volumes as supervision targets to finetune parameters in the forward model (not a reconstructed image). 

To study accelerated acquisition and reduced hardware cost, we evaluate (i) uniform subsampling over azimuthal scanning angles, (ii) element subsampling within each quarter-ring, and (iii) their combination. We report uniform acceleration at $6{\times}$ and $10{\times}$, and also assess limited-angle patterns in azimuth and elevation, using the same bowl geometry as the instrument.  

We consider different sensor subsampling settings (\cref{fig:teaser}b), which accelerate 3D-PACT or reduce the cost.
Specifically, we consider different subsampling patterns and subsampling/subsampling rates. For subsampling patterns, we considered full, limited angle in azimuth and elevation (bottom row of \cref{fig:teaser}b). This paper mainly considers uniform subsampling, as the original system is designed to have a rotating arc, and uniform subsampling directly has the physical meaning of accelerating the imaging (or the sensory data acquisition time). For the uniform subsampling, we consider different rates of acceleration. Specifically, top row of \cref{fig:teaser}b depicts the 6$\times$, 10$\times$ subsampling rate with uniform subsampling.

\begin{figure}[t!]
    \centering
    \includegraphics[width=\columnwidth]{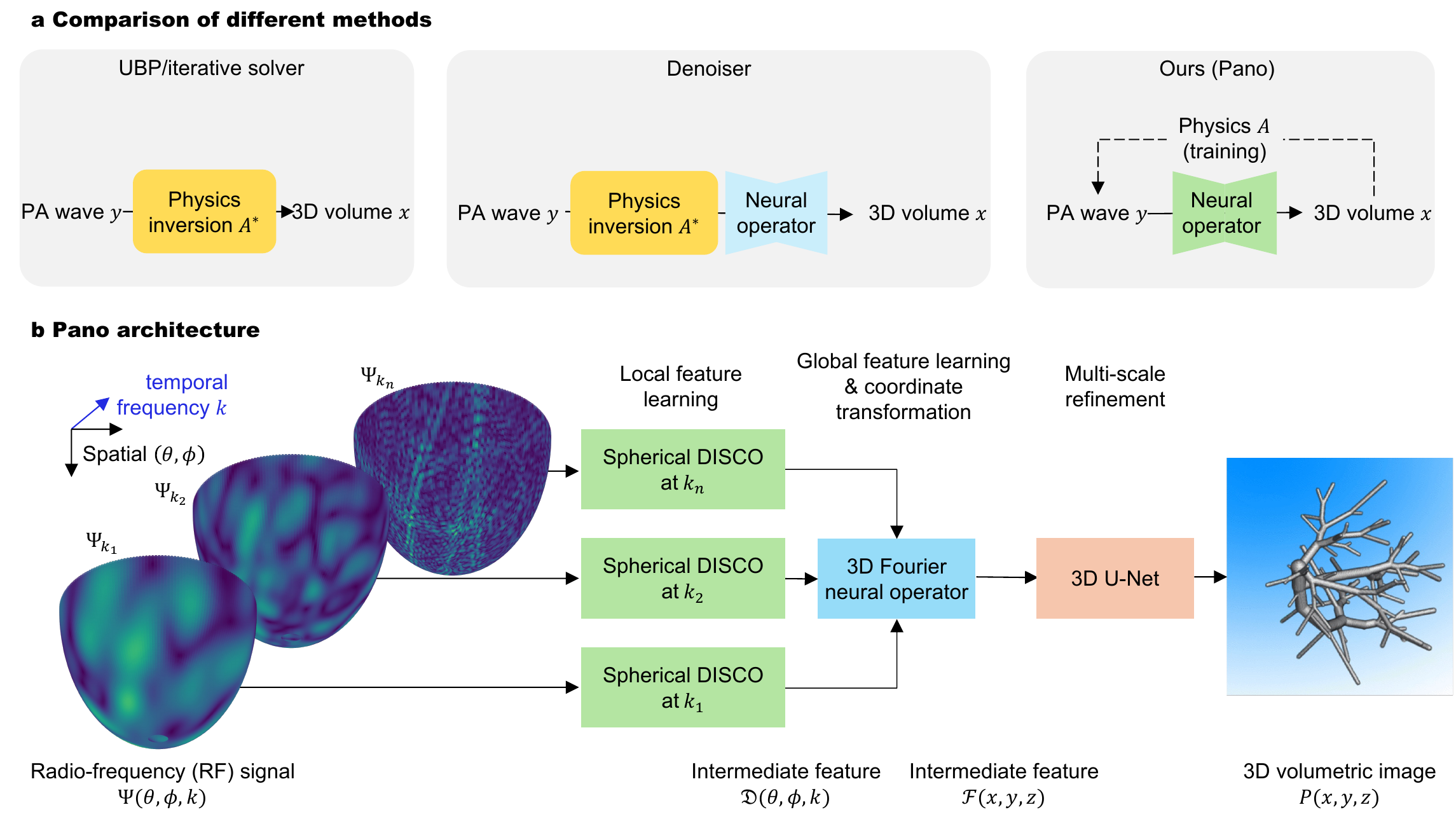} 
    \caption{\small {\bf \textsc{Pano} design and architecture.} 
    {\bf a, }Conceptual comparison of different methods. {\it 1)} Solver-based methods like UBP \cite{xu2005universal} directly invert the input $\Psi$ with a physical imaging model.  {\it 2)} Reconstruct-then-denoise method like \cite{choi2023DLPACT} first inverts the input $\Psi$ with the physics model and then uses a network (e.g. U-Net) to denoise/refine for better reconstruction. {\it 3)} The proposed \method is the first end-to-end method for 3D PACT reconstruction. It directly inverts $\Psi$ with a resolution-agnostic neural operator. \method is also physics-aware by enforcing the physical model during training. 
    {\bf b,} The design of the proposed \textsc{Pano} considers the physical model/sensing matrix $A$ of the imaging process $\Psi = AP$, where $A$ is the Helmholtz equation. Specifically, considering the Helmholtz equation is time-independent, different frequency $k_i$ components of the input PA wave $y_{k_i}$ are processed independently first with a local feature learning component, spherical DISCO (discrete-continuous convolution). Spherical DISCO is a neural operator block that mimics spherical convolution and makes \textsc{Pano} agnostic to different subsampling of the input measurement data (See \cref{fig:disco}). Multi-frequency features are then combined and fed to the global feature learning module, FNO (Fourier neural operator). FNO will also perform a coordinate transform: from spherical coordinates to Cartesian coordinates. Finally, a multi-scale feature learning module, 3D U-Net, outputs the reconstructed 3D volumetric image $P$.
    }
    \label{fig:fig2}
\end{figure}

\subsection{\method  for 3D PACT Reconstruction}
The overall architecture of the proposed deep learning framework, \method, for 3D PACT reconstruction is depicted in \cref{fig:teaser}c. \method is a neural operator that transforms the input PA (photoacoustic) wave $\Psi$ (sensory radiofrequency (RF) signal) to 3D volumetric image $P$. 
\method is a neural operator (NO), a deep neural architecture designed to learn maps between function spaces.  
The neural operator architecture is agnostic to the sampling pattern of the sensor array that detects the PA wave, making \method able to accelerate PA imaging from subsampled input measurements. \method reconstructs images at a fixed resolution due to the setting of the 3D PACT imaging. 
As a cycle consistency check to ensure physics validity, the reconstruction $\hat{P}$ is further projected back to PA waves, and a physics loss is used to penalize if the reconstruction's PA wave deviates significantly from the input $\Psi$. In the following, we briefly explain each component.

\bmhead{\method overview}
We illustrate the detailed architecture of \method in \cref{fig:fig2}b.
The input PA  (photoacoustic) wave is a radio-frequency signal in the temporal domain. We first apply the Fourier transform on the PA wave to make it in the frequency domain. 
Starting from the multi–frequency PA  wave
\(\Psi(\theta,\phi,k)\in\mathbb{R}^{N_\theta\times N_\phi\times N_k}\), our model \method seeks to recover the three-dimensional initial-pressure distribution  
\(P\in\mathbb{R}^{N_x\times N_y\times N_z}\). 
Specifically, the PA wave $\Psi$ is recorded in a hemispherical sensory array at the polar location \((\theta,\phi)\) and at different frequencies $k$. $x, y, z$ refer to the spatial coordinates of the output 3D volumetric image reconstruction.
The reconstruction is realized by a composite physics-aware deep learning architecture that we denote \(\mathcal{G}_{\Theta}\), whose functional form can be written compactly as  
\begin{align}
    \hat{P}=\mathcal{G}_{\Theta}(\Psi)
        =\mathcal{U}\!
           \left(
             \mathcal{F}
             \bigl(
               \mathrm{Concat}_{k}\,
               (\mathcal{D}_{k}(\Psi_{k}))
             \bigr)
           \right),
\end{align}
where the inner operator \(\mathcal{D}_{k}\) is a discrete–continuous convolution (DISCO) block \cite{jatyani2025unified,DISCO} acting on each individual frequency slice \(\Psi_{k}\),  
\(\mathcal{F}\) is a Fourier Neural Operator (FNO) \cite{li2020fourier} that couples the resulting feature maps across all frequencies and  
\(\mathcal{U}\) is a lightweight three-dimensional U-shaped neural network that further improves spatial detail for reconstruction performance. DISCO  processes local features of the input $\Psi$ with the local convolution design, whereas FNO aggregates and processes the global features. Empirically, we justify the design of \method with ablation studies (Section \ref{sec:simresults}).

\bmhead{Resolution-convergent operator learning}
\method can be considered as a \emph{neural operator} which learns mappings between \emph{function} spaces rather than finite-dimensional vectors.  
Such operators are \emph{resolution-convergent}: once trained at a given grid size they generalize seamlessly to unseen sensor samplings, whether uniformly subsampled, clustered or adaptively distributed.  
In practice this flexibility allows a single model to reconstruct 3D volumetric images $P$ from sparsely sampled or accelerated acquisitions, thereby reducing hardware cost and boosting frame rate without retraining.

\bmhead{Geometry-aware feature extraction}
Because the PA wave sensors lie on the surface of a hemisphere, we propose a DISCO block that performs learnable convolutions directly on the sphere \(S^{2}\).  
This spherical treatment preserves geodesic distances, eliminates the distortions inherent to planar projections (\cref{fig:disco}a), and endows the network with rotational equivariance, as illustrated in \cref{fig:fig2}b. We also align the axis of spatial coordinates of the PA wave $\Psi$ lying on the hemisphere and the target 3D reconstruction $P$ for the entire framework.

\bmhead{Global feature learning with FNO}
Frequency-specific features after DISCO block are first concatenated and then fed to a Fourier neural operator (FNO) \cite{li2020fourier}. FNO learns the global features spatially by perform the Fourier transform on the spatial coordinates of the signal in the sensory domain. The global feature learning block complements the DISCO, which is based on convolution, the local integral operator.

\bmhead{3D U-Net further refines the reconstruction}
We finally add a lightweight residual 3D U-Net \cite{ronneberger2015u,cciccek20163d} to further refine the image reconstruction, as 3D FNO learns in the low-frequency space and may not reconstruct high-frequency components of the image. Note that the U-Net works best when reconstructing images at a fixed resolution, which fits the 3D PACT imaging requirements, as there is no need for the flexibility to reconstruct images at different resolutions.

\bmhead{Physics-aware learning}
To anchor the network in physical validity we minimize a combined data and physics loss
\begin{align}
\mathcal{L}(\Theta)=
\lambda_{\mathrm{img}}\lVert \hat{P}-P\rVert_{1}
+
\lambda_{\mathrm{phys}}\lVert A\hat{P}-\Psi\rVert_{2}^{2},
\tag{2}
\end{align}
where \(A:\mathbb{R}^{N_x\times N_y\times N_z}\!\rightarrow\!\mathbb{R}^{N_\theta\times N_\phi\times N_k}\) is an operator solving the Helmholtz equation (the forward model of the PACT imaging system).  
The first term rewards voxel-wise fidelity in $P$, whereas the second projects the prediction back into measurement space and penalizes the sensory data PA-wave $\Psi$ mismatches.  
Because \(A\) is evaluated only during training, inference remains a single feed-forward pass with complexity \(O(|\Theta|)\). In other words, the physics loss only affects the training not the inference of the method. 
We also accelerate the training by randomly subsampling $A\hat{P}$ at different training steps.

To summarize, \method unites spherical DISCO for local feature learning, an FNO for global feature learning and a physics-aware loss during training. \method simultaneously respects detector geometry, adapts to arbitrary sampling patterns, and honors the governing wave equation.  
The resulting reconstruction of \method exhibits state-of-the-art quantitative accuracy while enabling faster, lower-cost data acquisition. 

\bmhead{Baseline methods} We mainly consider the following baselines, as depicted in \cref{fig:fig2}a: {\bf 1)} Solver-based methods. Such methods rely on the physical model of the imaging and are thus learning-free. We consider UBP (universal back-projection algorithm) \cite{xu2005universal} and iterative solver \cite{xu2002exact,zhang2009effects}.  {\bf 2)} Learning-based method. We follow DL-PACT \cite{choi2023DLPACT} for the reconstruction-and-denoising framework. We refer to such a method as a ``denoiser'' as it is not an end-to-end method but denoises the physics solver reconstruction. 
On the setting of real data with $15\times$ subsampling, \cref{fig:teaser}d compares the performance (cosine similarity) and inference speed of different reconstruction algorithms. \method achieves improved reconstruction performance with faster or similar inference time over the baseline methods. 






\subsection{In~Silico Results}
\label{sec:simresults}
\bmhead{Synthetic data generation}
We generated paired \emph{volume–RF} samples $(\mathbf{x},\mathbf{y})$ by (i) synthesizing 3-D vascular phantoms with \textit{VascuSynth} \cite{cmig2010} (see \textit{Initial pressure generation}), (ii) mapping them to initial pressure fields $P(\mathbf{r})$ (proportional to optical absorption; positivity enforced), and (iii) propagating $P$ with the homogeneous acoustic model to simulated transducer data (see Section \ref{sec:forwardm} \textit{Physical forward model}). Briefly, VascuSynth produced binary vessel volumes on a grid, which we resampled to $200{\times}200{\times}160$ voxels (voxel pitch $\Delta x=0.25\text{\,\;mm}$), smoothed mildly to avoid staircase artifacts, and scaled to a nominal peak pressure to set SNR. The acoustic forward operator included the measured/effective receive impulse response, band-limit to the transducer bandwidth, and DFT readout over $N_f=149$ positive frequencies. Each channel’s time trace was windowed to $T=1000 \;\text{ms}$ and sampled at $f_s=10\; \text{MHz}$.

\underline{Realism and anti–inverse-crime.} To better match experiments while avoiding model overfit, we added: (i) per-scan speed-of-sound jitter 
(ii) per-channel gain and timing offsets (calibrated/whitened as in the Methods); and (iii) complex Gaussian noise to reach target $\mathrm{SNR}\in (1, 10)\; \text{db}$. Pre- and post-processing (baseline removal, band-pass, time-zero alignment) followed the same settings used for reconstruction (see \textit{UBP} and \textit{Iterative reconstruction}).

\underline{Measurement sparsity.} For each synthesized volume we created challenged acquisitions to study ill-posedness:
\begin{itemize}
  \item \textbf{Uniform down-sampling:} retain every $k$-th detector ($k\in\{6,10,15,20\}$), yielding $6{\times}$–$20{\times}$ sub-Nyquist sampling.
  \item \textbf{Limited angle:} restrict detectors to a $120^\circ$ azimuthal arc (or keep only the most proximal ${\sim}\tfrac{1}{3}$ of sensors near the source), inducing pronounced limited-view artifacts.
\end{itemize}

\bmhead{Results}
\begin{figure}[t!]
    \centering
    \includegraphics[width=\columnwidth]{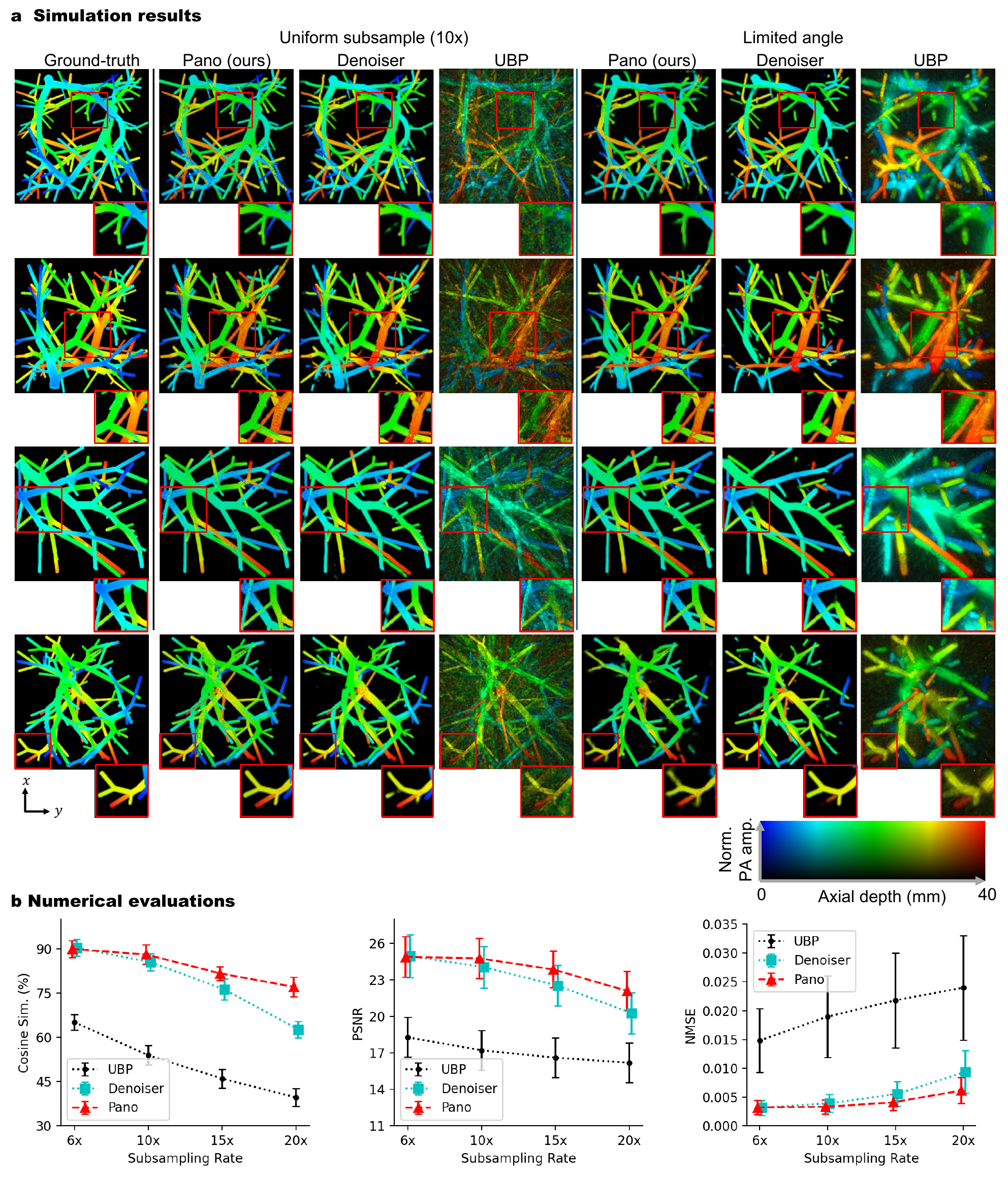} 
    \caption{\small {\bf Results on simulated data.}
    {\bf a, } Visualization of 3D reconstruction of different methods (the proposed method \textsc{Pano}, Denoiser \cite{choi2023DLPACT} and UBP (universal back-projection) \cite{xu2005universal}. We visualize with MAP (maximum amplitude projection on the z-axis) and observe that the proposed \method reconstructs 3D images with high-fidelity.  
    Zoomed-in view is provided on the bottom right of each subfigure for easier visualization. We consider both the uniform subsampling setting and the limited-angle reconstruction ($\frac{1}{3}$ in full elevation) setting, as shown in \cref{fig:teaser}b. We use HSV color space as the color coding, where the axial depth is encoded as hue, while the normalized PA (photoacoustic) amplitude is encoded as value. 
    {\bf b, } Quantitative evaluation (cosine similarity, PSNR, NMSE) across subsampling rates $6{\times}$--$20{\times}$. \method matches the Denoiser at $6{\times}$ ($-$0.4 percentage points) and progressively outperforms it at higher acceleration, reaching a 14.4 percentage point advantage at $20{\times}$, confirming that end-to-end operator learning degrades more gracefully than the two-step paradigm under aggressive subsampling.
    }
    \label{fig:sim_results}
\end{figure}

\begin{figure}[t!]
    \centering
    \includegraphics[width=\columnwidth]{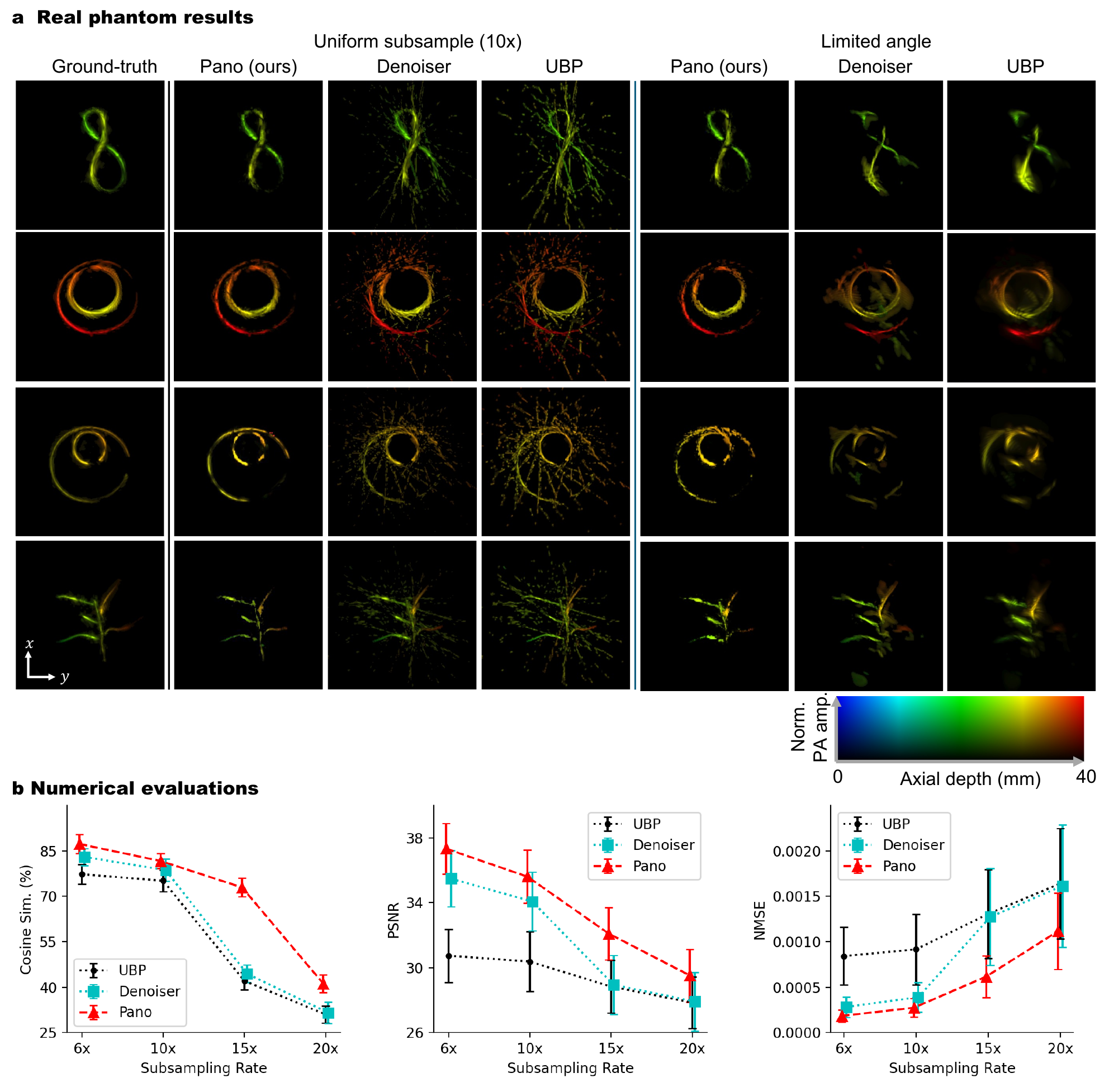} 
    \caption{\small {\bf Results on real data.}
        {\bf a, } Visualization of 3D reconstruction of different methods (PA MAP images of the proposed \method, Denoiser \cite{choi2023DLPACT} and UBP (universal back-projection) \cite{xu2005universal}. We consider both the uniform subsampling setting and the limited-angle reconstruction ($\frac{1}{3}$ in full elevation) setting, as shown in \cref{fig:teaser}b. Compared to other methods, \method reconstruct 3D images with improved fidelity-persevering details and less noise. We use the HSV color space, where the axial depth is encoded as hue while the normalized PA (photoacoustic) amplitude is encoded as value. 
    {\bf b, } Quantitative evaluation (cosine similarity, PSNR, NMSE) on real phantom data. \method outperforms UBP and the Denoiser at all four subsampling rates, with the largest margin at $15{\times}$ (28.4 percentage points over Denoiser), validating sim-to-real generalizability of the proposed approach.
    }
    \label{fig:real_results}
\end{figure}

\underline{Performance under different subsampling rates.} 
\cref{fig:sim_results}a qualitatively compares 3D reconstructions obtained with our neural operator \method,  
a U-Net denoiser, and the analytic universal back-projection (UBP) algorithm. 
For visualization purposes, we provide 2D PA MAP (maximum amplitude projection on the z-axis) images and use color encoding for PA amplitude and axial depth (along the z-axis). The 3D volumetric image has a resolution of $200\times200\times160$ and the PA image has a resolution of $200\times200$. 
Under \emph{uniform  
$10\times$ down-sampling}  (left of \cref{fig:sim_results}a), our method \method faithfully preserves details of the 3D vessel structures, whereas the denoiser misses fine branches and UBP exhibits pronounced streak artifacts.  
The performance gap widens under the \emph{limited-angle} setting (right of \cref{fig:sim_results}a), where \method maintains coherent vessel  
topology while competing methods collapse into depth-dependent noise.

The numerical performance is summarized in \cref{fig:sim_results}b and Supplementary Table~\ref{tab:sim_results}. \method substantially outperforms the UBP baseline: over 6$\times$, 10$\times$, 15$\times$, and 20$\times$ acceleration, cosine similarity improves by 25, 34, 36, and 37.4 percentage points, respectively, yielding an average gain of approximately 33 percentage points. The average PSNR gain over UBP is 6.8\,dB, and the average NMSE reduction is 0.0157.
Compared to the deep learning baseline~\cite{choi2023DLPACT}, \method's cosine similarity improves by 0, 3, 5, and 14 percentage points at the same four subsampling rates. While the two methods perform comparably at low acceleration ($6\times$), the gap widens markedly as measurements become sparser. In terms of PSNR, the average gain over the Denoiser across all four rates is 1.0\,dB, and the average NMSE reduction is 0.0023.
These trends indicate that the two-step Denoiser degrades rapidly under high acceleration, whereas \method maintains robust reconstruction quality by directly inverting the forward operator. The gains at $20\times$ underscore the advantage of end-to-end operator learning for highly subsampled acquisitions.

\begin{figure}[t!]
    \centering
    \includegraphics[width=\columnwidth]{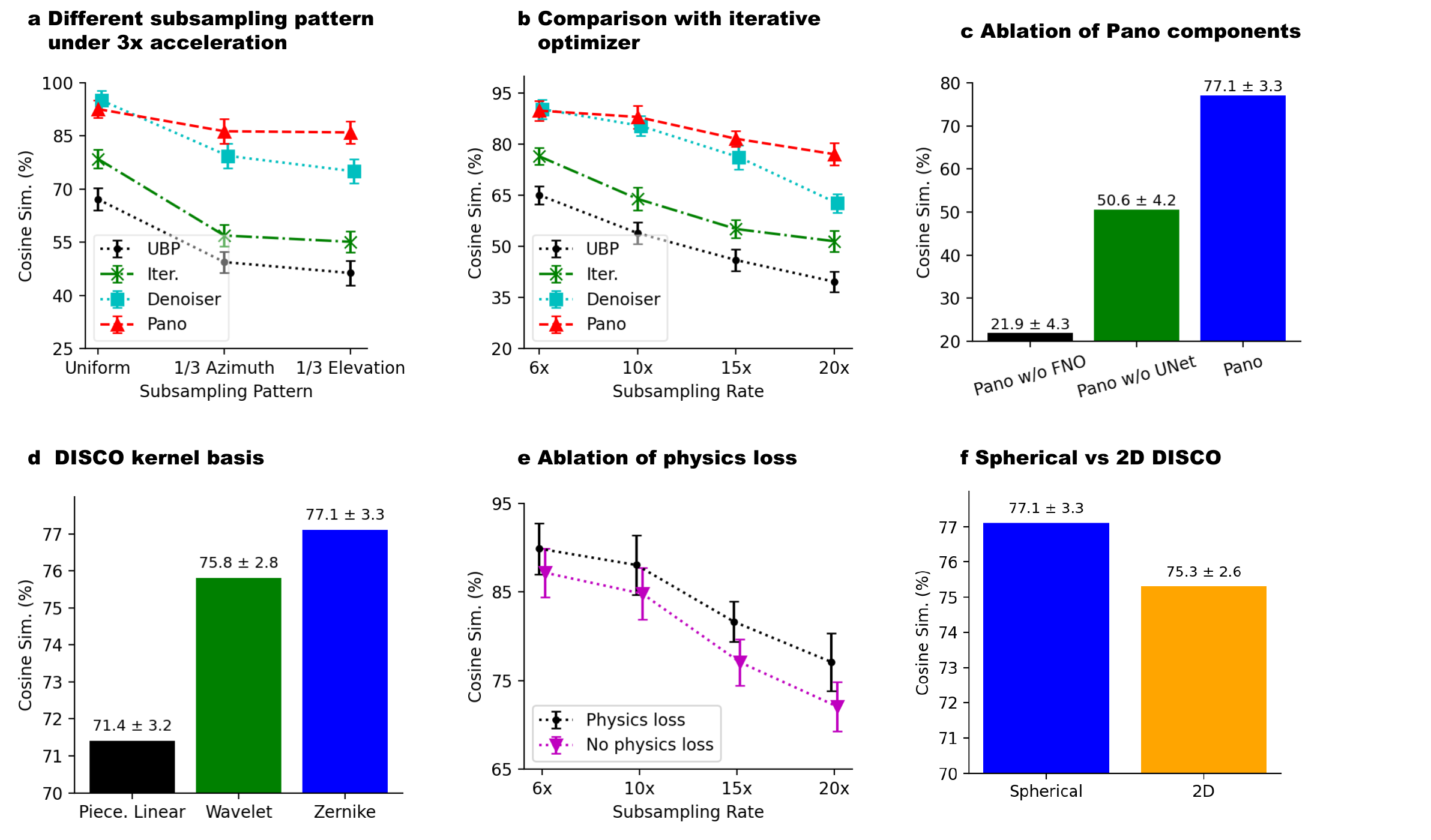} 
    \caption{\small {\bf Analysis and ablation study.}
            {\bf a, } All methods' performance with different subsampling patterns under 3$\times$ acceleration. \method outperforms baselines on limited azimuth and elevation settings, and is on par with Denoiser \cite{choi2023DLPACT} under the uniform sampling setting. 
    {\bf b, } Comparison with iterative solvers \cite{xu2002exact} on the simulated data. With uniform subsampling, we observe an improved performance of the iterative solver over UBP \cite{xu2005universal}, at the cost of approximately 10$\times$ slower in inference time.
    {\bf c, } Ablation study of different \textsc{Pano} components. Removing the FNO block leads to a dramatic performance drop (55.2\%), indicating the power of the global feature learning of the FNO.
        {\bf d, } Performance under different DISCO kernel bases at $20{\times}$ subsampling. The Zernike basis achieves the best cosine similarity (77.1\%), followed by wavelet and piecewise linear (71.4\%).
    {\bf e, } Ablation study of the physics loss. Adding the physics loss improves PACT reconstruction performance under different subsampling rates.
    {\bf f,} Comparing spherical versus 2D DISCO \cite{DISCO}. 2D DISCO directly projects the spherical coordinates into a Cartesian grid and leads to 2\% performance drop compared to spherical DISCO used in the proposed \method.
    }
    \label{fig:ablation}
\end{figure}

\underline{Performance under different subsampling patterns.} 
We report the performance of different methods under different subsampling patterns (uniform, limited angle in azimuth and limited angle in elevation, as depicted in \cref{fig:teaser}b) under the same subsampling rate 3$\times$ (\cref{fig:ablation}a). Averaged across all patterns, UBP and the iterative solver achieve a cosine similarity of 54.3\% and 63.5\%. Denoiser and the proposed \method achieve performance of 83.2\% and 88.3\%, respectively. The proposed \method thus has 5\% gain over a learning-based method and 25\% gain over the iterative solver.  Detailed numerical results are provided in Supplementary Table~\ref{tab:dpattern}.

\underline{{Comparison with iterative solvers.}}  The comparison with the iterative solver can be found in \cref{fig:ablation}b. With uniform subsampling, we observe an improved performance of the iterative optimizer over UBP, at the cost of approximately 10$\times$ slower in inference time. Note the estimation is on 5 iterations of running the iterative solver, with which setting we empirically obtain a converged result with metric of cosine similarity of reconstruction and the ground truth.



Overall, the in-silico study demonstrates that our \method delivers improved reconstruction accuracy over existing state-of-the-art methods  
across severe sub-sampling and limited-angle scenarios, while affording orders-of-magnitude faster inference time than conventional solvers and outperforming representative deep-learning baselines. We report the ablation study results of the proposed method to justify our design choice.

\underline{Ablation of major \method components.} \method contains three major components: a DISCO block that processes measurement data in a resolution-agnostic way, a 3D FNO block that learns global features and performs coordinate transforms, and finally a 3D U-Net to further refine the image reconstruction.
We study the functions of FNO and U-Net by removing them in \method and report the results in \cref{fig:ablation}c. Removing U-Net leads to a 26.5 percentage point performance drop. Visual results show that while global structures are still retained, removing U-Net leads to more missing of local 3D structures. Removing FNO leads to a drastic 55.2 percentage point performance drop, where the majority of the global 3D structures are not retained. This indicates that it is necessary to use a global feature learning and coordinate transform (from spherical coordinates to Cartesian coordinates) block for 3D PACT reconstruction.

\begin{figure}[t!]
    \centering
    \includegraphics[width=\columnwidth]{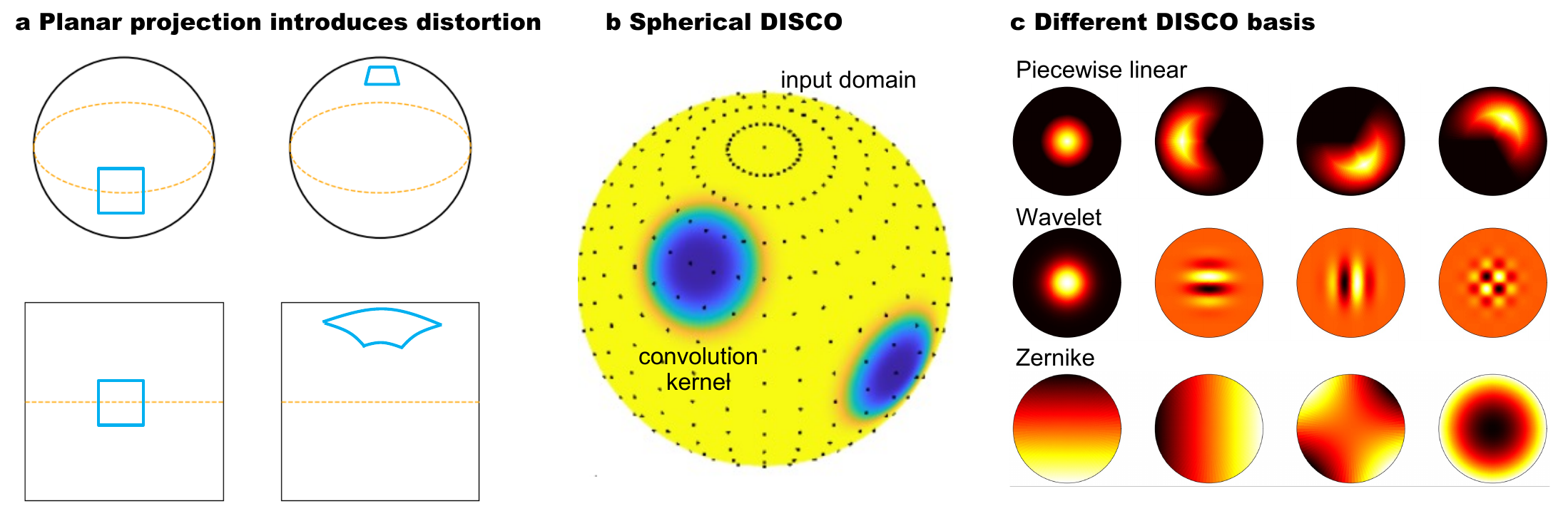} 
    \caption{\small {\bf Spherical DISCO design details.} 
    {\bf a,} Motivation for using spherical convolution: Planar projection of a spherical signal will result in distortions. Rotation of a spherical signal cannot be emulated by translation of its planar projection. In the figure, with Mercator projection, an equatorial patch (left) stays compact, while the same-area patch near the pole (right) inflates heavily in the vertical direction.
    {\bf b,} We use the spherical convolution parameterized with DISCO kernels to process PA wave sampled at hemispherical sensors, achieving distortion-free learning with a resolution-agnostic design.
    {\bf c, } Different DISCO kernel bases considered in the work. See Section \ref{sec:panodesign} for details.
    }
    \label{fig:disco}
\end{figure}

\underline{Ablation of DISCO kernels.} 
We consider three different DISCO kernels, piece-wise linear, wavelet and Zernike, with details in Section \ref{sec:panodesign}. We visualize the three kernel configurations in \cref{fig:disco}c. 
\cref{fig:ablation}d depicts the performance under different DISCO \cite{DISCO} kernels. The setting is 20$\times$ uniform subsampling. We observe that the Zernike basis gives the best performance at 77.1\%, while the piecewise linear basis gives the worst performance at 71.4\%. Other than specifically mentioned, we use the Zernike basis for \method when reporting the numerical performance.

\underline{Ablation of physics loss.} \cref{fig:ablation}e depicts the training convergence and performance of the physics loss. The numerical performance comparison can be found in Supplementary Table~\ref{tab:physics_loss}. With physics loss, we observe an average gain of 3.9\% for different subsampling rates under the uniform subsampling pattern. 

\underline{Spherical vs 2D DISCO} Instead of using spherical DISCO \cite{DISCO}, we also compare with 2D DISCO setting. Specifically, instead of performing spherical convolution, we first project the sensory data from the spherical coordinate domain to the 2D Cartesian domain, and then perform the 2D DISCO (as defined in \cite{jatyani2025unified}) on the projected domain. The number of parameters is kept the same and the output shape is the same as the spherical counterpart. Note that such 2D projection and convolution would lead to distortion as shown in \cref{fig:teaser}b. 
In \cref{fig:ablation}f, we report a 2\% performance drop (in cosine similarity) when using the 2D projected distorted convolution.

\subsection{Real Experimental Data Results}

To evaluate the generalizability of \method for the experimental data, we acquired dense PA measurements of phantoms made of black wires. A densely sampled scan ($k\!=\!1$) serves as a proxy ground-truth volume, while subsets of the raw channel data were retrospectively down-sampled to yield the uniform subsampling rate $\in\!\{6,10,15,20\}$ and {$120^{\circ}$ limited-angle} regimes used for testing. A small calibration set of $N_{\!ft}=37$ point sources (see Methods) was employed for two-stage fine-tuning of the proposed \method; Denoiser baselines were fine-tuned identically. Details of the finetuning can be found in Section \ref{sec:implementation}.

\bmhead{Results}

\underline{Qualitative comparison.}
\cref{fig:real_results}a shows representative reconstructions.
In the \textbf{uniform $10\times$} case, \method cleanly reconstructs  3D phantom structures (e.g. loop and ring), whereas Denoiser blurs thin segments and
UBP introduces characteristic radial streaks.  
Under the \textbf{limited-angle} setting the advantage becomes more pronounced:
\method retains coherent morphology,  while existing methods collapse into patchy artifacts or depth misregistrations.
The qualitative fidelity mirrors the simulated study, confirming that physics-aware operator learning generalizes to experimental imperfections.

\underline{Quantitative evaluation}.
The numerical performance is summarized in \cref{fig:real_results}b and Supplementary Table~\ref{tab:real_results}. Metrics are averaged over 14 real phantom test volumes at each subsampling rate. \method outperforms the UBP baseline by a large margin: cosine similarity improves by 10, 7, 31, and 10 percentage points at 6$\times$, 10$\times$, 15$\times$, and 20$\times$ acceleration, respectively, for an average improvement of over 14 percentage points. The average PSNR gain over UBP is 4.2\,dB, and the average NMSE reduction is 0.0006.
Compared to the deep learning baseline~\cite{choi2023DLPACT}, \method improves cosine similarity by 4, 3, 28, and 10 percentage points at the same four rates, with an average gain of 11 percentage points. The most striking improvement occurs at $15\times$, where the Denoiser struggles with severe sparsity while \method remains robust. The average PSNR gain over the Denoiser is 2.0\,dB, and the average NMSE reduction is 0.0003.
Taken together, the real-data results mirror the in-silico findings, confirming that direct inverse operator learning provides consistent and growing advantages over two-step methods as acquisition becomes more challenging.

\underline{Runtime.}  
Thanks to a single forward pass of a lightweight network, \method reconstructs
a $200\times200\times160$ volume in \textbf{0.11\,s} on an NVIDIA RTX 4090 GPU,
corresponding to an {effective 9 Hz 3D display rate}.
This real-time capability is pivotal for interactive PA imaging. 
On the setting of real data with $15\times$ subsampling, \cref{fig:teaser}d compares the performance (cosine similarity) and inference speed of different reconstruction algorithms. \method achieves improved reconstruction performance with similar speed as UBP.
The 0.11\,s inference time represents a substantial step toward interactive 3D PACT; further reductions may be achievable through model compression and hardware-optimized inference pipelines.

The close correspondence between simulated and experimental metrics indicates that (i) the domain gap introduced by acoustic heterogeneity and system noise is modest, and (ii) limited fine-tuning suffices to bridge it. 
Together, these results establish our \method as a robust and computationally efficient solution for practical 3D photoacoustic tomography. The strong simulation-to-real generalization of \method motivates future in-vivo studies as a step toward broader clinical applicability.


\section{Discussion}

We propose \method, a first end-to-end 3D PACT image reconstruction method that departs fundamentally from the prevailing \emph{reconstruction-and-denoising} paradigm.  
Classical 3D PACT pipelines center on physics-based solvers—most commonly universal back-projection (UBP)~\cite{xu2005universal}—to reconstruct an initial volume from measured PA wave.
Some follow a \emph{reconstruction-and-denoising} pipeline \cite{choi2023DLPACT,cao2023single} and further add a neural network after the solver to remove image artifacts. Because the network never \emph{sees} the measurements, performance and robustness hinge on the solver’s fidelity and per-case hyperparameter tuning. In contrast, \method is, to our knowledge, the first end-to-end deep learning framework that (i) \emph{directly} maps raw PA wave data to a high-fidelity 3D volume and (ii) \emph{jointly} learns a physics operator and a data prior—thereby performing the physical inversion while refining reconstructions using priors learned from the training distribution. By embedding the imaging physics (the partial differential equation that models the forward imaging process) in the learning process, \method maintains consistency with the forward model while leveraging statistical regularities in anatomical structure. This leads to two practical advantages: \textbf{ 1)}{ Generalization.} The model remains reliable when sampling density, noise statistics, or targets deviate from training, as evidenced by consistent performance across simulated and real experiments.  \textbf{ 2)}{  Simplified deployment.} Unlike UBP \cite{xu2005universal}, which requires per-sample solver hyperparameter tuning, \method’s hyperparameters are fixed at training time and transfer across samples and acquisition settings.

Three architectural ingredients work in concert to enable end-to-end, physics-aware reconstruction. First, \method operates natively on the hemispherical sensor manifold as a neural operator, remaining agnostic to input measurement resolution and accommodating different subsampling patterns within the same sensor array geometry without retraining—eliminating the need for model variants tied to specific subsampling patterns. Together with a $0.11$\,s inference latency for a $160\times200\times200$ volume on a single NVIDIA RTX 4090 GPU, this design supports real-time 3D PACT reconstruction and visualization. Second, DISCO~\cite{DISCO} kernels perform convolutions directly on the sphere, preserving the geodesic locality that planar kernels distort and allowing \method to exploit the symmetries of the hemispherical sensor. Third, the architecture is informed by the Helmholtz equation: a differentiable wave-propagation loss on the reconstruction $\hat{P}$ enforces acoustic consistency to mitigate hallucinations. This tight coupling of geometry and physics distinguishes \method from purely data-driven denoisers~\cite{choi2023DLPACT} and underpins its resilience to varied sampling schemes.

We attribute the observed performance gains to both a fundamental reframing of the problem and these three technical design choices. By replacing the reconstruct-then-refine paradigm with direct inverse operator learning, \method combines physics and data priors in a single pass and is inherently resolution-agnostic within the same sensor array geometry. Spherical DISCO preserves neighborhood structure on the hemispherical detector manifold; FNO captures the global interactions required by the inverse mapping; and physics consistency regularizes predictions to remain physically plausible. Ablation experiments confirm that removing any one of these components causes substantial degradation, with the effect becoming especially pronounced at high acceleration factors.

Although we validate only on PACT, the underlying principle---learning inverse operators while constraining them with forward physics---applies broadly to other inverse problems with expensive forward models, including ultrasound tomography, seismic imaging, radar imaging, and optical tomography. We present this as a methodological implication, not as an experimentally validated cross-domain claim.

Despite these strengths, \method has some limitations. The training still requires substantial GPU memory (ideally an 80 GB NVIDIA A100 GPU) because both the input PA wave and the output reconstruction are three-dimensional and have relatively high resolution. 
Future work will explore implicit neural representations, such as point clouds \cite{wang20223d}, signed-distance fields \cite{park2019deepsdf} or neural fields \cite{mildenhall2021nerf}, to compress the voxel grid and push to higher resolutions. Such efficient 3D representation may be beneficial for making \method a generative model, which has been shown useful in recent literature for 2D inverse problems \cite{cao2025diff,zheng2025inversebench}. 
Additionally, the current formulation relies on two significant physical simplifications: a spatially homogeneous sound speed and a uniform optical fluence ($\Phi(\mathbf{r}) = \Phi_0$).  In real tissue, optical fluence is highly depth- and wavelength-dependent, which means the recovered initial pressure field conflates true optical absorption with fluence variations.  Extending \method to heterogeneous acoustic media and incorporating a fluence correction model---for example, via a coupled diffusion--wave solver---will be essential for quantitatively accurate transcranial or abdominal photoacoustic imaging.  On the experimental side, \emph{in-vivo} animal and human studies are warranted to further validate the proposed \method, as the current experiments are on phantoms. 
Moving from phantoms to living tissue will likely require several targeted modifications to \method, including heterogeneity-aware acoustics design, motion robustness, domain adaptation with limited data, and uncertainty and out-of-domain detection.
Specifically, densely sampled ground truth for in-vivo imaging is impractical to acquire in living tissue, which makes rigorous benchmarking challenging. 

In summary, by recasting photoacoustic reconstruction as an operator-learning problem that integrates physics and data priors, \method delivers high-quality, real-time 3D imaging from sparse measurements, eliminates solver tuning and scales seamlessly across resolutions.  These attributes pave the way for compact, cost-effective PACT systems suitable for preclinical research, and motivate future in-vivo studies as a step toward eventual bedside applications.

\section{Materials and Methods}
\subsection{Inverse-operator learning formulation}
We cast 3D PACT reconstruction as learning a direct inverse map from measured RF data to initial-pressure volume,
\begin{align}
\hat{P}=\mathcal{G}_{\Theta}(\Psi),
\end{align}
where $\mathcal{G}_{\Theta}$ is a neural operator parameterized by $\Theta$, $\Psi$ is the multi-channel measurement in frequency domain, and $\hat{P}$ is the reconstructed volumetric initial-pressure image.
To preserve physical fidelity, training uses both image-domain supervision and a forward-consistency constraint through the acoustic operator $A$:
\begin{align}
\mathcal{L}=\lambda_{\mathrm{img}}\,\mathcal{L}_{\mathrm{img}}(\hat{P},P)+\lambda_{\mathrm{phys}}\,\mathcal{L}_{\mathrm{phys}}(A\hat{P},\Psi).
\end{align}
This formulation keeps the reconstruction end-to-end while retaining explicit links to the governing forward physics.

\subsection{Physical  Model}
We describe the physical model used in the paper, including the forward PA wave propagation model and the inverse problem of initial pressure $P(\mathbf{r}) $ (3D volumetric image) reconstruction.

\bmhead{Forward model}
\label{sec:forwardm}
\underline{Initial pressure.} A short laser pulse deposits optical energy that thermoelastically launches an initial pressure rise
\begin{align}
    P(\mathbf{r}) \;=\; \Gamma(\mathbf{r})\,\mu_a(\mathbf{r})\,\Phi(\mathbf{r}),
\end{align}
where $\Gamma$ is the Grüneisen parameter, $\mu_a$ is the optical absorption coefficient, and $\Phi$ is the optical fluence. Our imaging objective is to recover $P(\mathbf{r})\ge 0$.

\underline{Acoustic propagation.} The pressure field $p(\mathbf{r},t)$ evolves according to the acoustic wave equation
\begin{align}
\frac{1}{c^2(\mathbf{r})}\,\frac{\partial^2 p}{\partial t^2}(\mathbf{r},t)\;-\;\nabla\!\cdot\!\left(\frac{1}{\rho(\mathbf{r})}\nabla p(\mathbf{r},t)\right)\;=\;0,
\quad
p(\mathbf{r},0)=P(\mathbf{r}),\quad
\partial_t p(\mathbf{r},0)=0,
\end{align}
with $c(\mathbf{r})$ and $\rho(\mathbf{r})$ the sound speed and density. In this work we adopt the standard homogeneous model for PACT reconstruction, $c(\mathbf{r})\equiv c_0$ and $\rho(\mathbf{r})\equiv\rho_0$, and impose absorbing boundaries (PML) in numerical solvers. Frequency-dependent attenuation, when modeled, is incorporated by a complex wavenumber $k(\omega)=\omega/c_0 + i\,\alpha(\omega)$ (power-law $\alpha$).

\underline{Sensing model.} Let $\{\mathbf{s}_m\}_{m=1}^{N_d}$ denote detector positions. Each channel records a band-limited, impulse-convolved version of the pressure at the aperture plus noise:
\begin{align}
y_m(t) \;=\; \big(h_{\mathrm{rx}} * p(\mathbf{s}_m,\cdot)\big)(t) \;+\; \eta_m(t),
\end{align}
where $h_{\mathrm{rx}}$ is the receive (and electronics) impulse response, and $\eta_m$ models measurement noise. After windowing the time traces to $T$ and applying a discrete Fourier transform (DFT) we retain $N_f$ positive-frequency bins, yielding complex spectra
\begin{align}
\Psi_{m,k} \;=\; H_{\mathrm{rx}}(\omega_k)\, \widehat{p}(\mathbf{s}_m,\omega_k) \;+\; \eta_{m,k},
\qquad \omega_k = \tfrac{2\pi k}{T},\quad k=1,\ldots,N_f .
\end{align}

Under the homogeneous model, $\widehat{p}$ admits the single-layer potential form
\begin{align}
\widehat{p}(\mathbf{s}_m,\omega)
\;=\;
\int_{\Omega}\! P(\mathbf{r})\,
\frac{e^{i\,k(\omega)\,\|\mathbf{r}-\mathbf{s}_m\|}}{4\pi\,\|\mathbf{r}-\mathbf{s}_m\|}\, d\mathbf{r},
\end{align}
i.e., a convolution with the free-space Green’s function.

\underline{Discretizations.} We discretize the volume on a Cartesian grid with $N = N_x N_y N_z = 200\times200\times160$ voxels and stack $P(\mathbf{r})$ into a vector $P\in\mathbb{R}^N$. Likewise, we stack the complex spectra from all detectors and retained frequency bins into $\Psi\in\mathbb{C}^M$ with $M=N_d N_f$ (here $N_f=149$). The forward operator $A:\mathbb{R}^N\!\to\!\mathbb{C}^M$ factors as
$A \;=\; S\,\mathcal{F}_t\,\mathcal{G}\,$
where $\mathcal{G}$ maps $P$ to time-domain pressures at all detector locations (time-domain solver or frequency-domain Green’s integral with $c_0$), $\mathcal{F}_t$ is the temporal DFT restricted to the $N_f$ positive frequencies and multiplied by $H_{\mathrm{rx}}(\omega)$, and $S$ stacks channels and applies per-detector quadrature/solid-angle weights if needed.

\bmhead{Inverse problem}
We recover $P$ by solving a noise-aware, regularized least squares problem with nonnegativity:
\begin{align}
\hat{P}
\;=\;
\underset{P\ge 0}{\arg\min}\;
\frac{1}{2}\,\big\|\, W\,(A P - \Psi)\,\big\|_2^2 \;+\; \mathcal{R}(P).
\end{align}
Here $W\succeq 0$ whitens the residuals (e.g., $W=\Sigma_\eta^{-1/2}$ from noise calibration across channels; $W=I$ if unknown), and $\mathcal{R}$ promotes physically plausible images (e.g., isotropic total variation (TV) or TV+$\ell_2$). All norms on complex vectors use $\|z\|_2^2=\sum_i |z_i|^2$. The positivity constraint reflects $P(\mathbf{r})\ge 0$ for nonnegative absorbed energy and $\Gamma>0$.


\subsection{Initial pressure generation}
We synthesized vascular phantoms using \textit{VascuSynth} \cite{cmig2010}, 
an ITK-based tool that procedurally grows 3-D vascular trees under hemodynamic and perfusion constraints and exports a volumetric image and the corresponding topology. For each phantom, we specified (i) a perfusion/root point, (ii) the target number of terminal nodes (the number of nodes is uniformly sampled from the range of [120, 500] leaves each time), and (iii) physical/growth parameters (e.g., \texttt{PERF\_FLOW}=8.33, viscosity \texttt{RHO}=0.036, bifurcation exponents \texttt{LAMBDA}=2.0, \texttt{MU}=1.0, asymmetry \texttt{GAMMA}=3.0). The voxel width for the synthesized volume was set to $\Delta x_\mathrm{syn}=0.25\; \text{mm}$ via the \texttt{voxelWidth} argument. We used random seeds to generate a cohort of anatomically varied trees.

To control spatial distribution, we provided a piecewise-constant oxygenation (demand) map that concentrated growth within a user-defined box 
and excluded regions outside it; the supply map was kept uniform unless noted. VascuSynth produced a stack of 2-D slices (volumetric image) and a GXL file describing the tree geometry; we ignored the optional image-degradation/noise settings to keep the initial pressure strictly ground-truth.

The resulting binary vessel volume $V(\mathbf{r})\in\{0,1\}$ was resampled (cubic) to our simulation grid ($200\times200\times160$ voxels; voxel pitch $\Delta x=0.25\; \text{mm}$) and lightly smoothed (Gaussian, $\sigma=1\;\text{vox}$) to avoid staircase artifacts at the acoustic grid scale. 
We then defined the initial pressure as
\begin{align}
P(\mathbf{r}) \;=\; P_0 \,\big(V(\mathbf{r}) * g_\sigma(\mathbf{r})\big),
\end{align}
with $P_0=133,000\;\text{mmHg}$ a global scaling parameter chosen to match the target SNR and $g_\sigma$ a mild low-pass kernel to keep spectral content within the acoustic bandwidth ($f_{\max}=3\;\text{MHz}$). When reporting vessel dimensions, we measured centerline-based radii on the synthesized geometry and confirmed that the resulting radius range was $r_x\in [-25, 25]$ mm, $r_y\in [-25, 25]$ mm , $r_z\in [10, 50]$ mm. These $P(\mathbf{r})$ fields served as the initial conditions for the forward acoustic model to generate simulated PA data.

\subsection{Imaging System}
The imaging system uses a custom 1024-element ultrasonic array arranged as four 256-element quarter-rings on a hemispherical bowl, one-to-one mapped low-noise preamplifiers and multi-channel DAQ (data acquisition system), and an azimuthal scanning mechanism. Each element has a 1.5$\times1.5 \text{m}^2$ active area with 2.4 mm pitch, 2.12 MHz center frequency, and one-way $6$dB bandwidth of 1.73 MHz (78\% fractional). Signals are digitized at 20 MHz (12-bit) with a 7.5 MHz analog anti-alias filter. The quarter-rings are mounted on a 26 cm-diameter PTFE hemispherical bowl filled with deionized water as the coupling medium; an engineered diffuser expands the beams to $\sim$10 cm on the phantom. We used dense scans (400 azimuthal angles) to generate the reference image.
For the point-source data, we couple 532 nm light from a laser (IS8-2-L, Edgewave) to an optical fiber (FG050LGA, Thorlabs; core diameter: $50\,\mu\mathrm{m}$) terminated with a light-absorbing material (carbon nanopowder), which acts as a point source for PACT. For the phantom data, a 1064 nm Nd:YAG laser is used to illuminate the black wire phantom, which generates the photoacoustic signal measured by the transducers.

\subsection{Image Reconstruction Baselines}

\bmhead{Universal back-projection (UBP)}
We reconstructed all images using the universal back-projection (UBP) algorithm \cite{xu2005universal}. In brief, UBP inverts the spherical Radon transform under the assumption of a spatially homogeneous acoustic speed $c_0$ and point-like detectors distributed on a measurement surface $\mathcal{S}$. For a voxel at $\mathbf{r}$ and detector at $\mathbf{s}\in\mathcal{S}$ with line-of-sight distance $R(\mathbf{r},\mathbf{s})=\|\mathbf{r}-\mathbf{s}\|$, the reconstruction evaluates the time-of-flight $t^\star=R(\mathbf{r},\mathbf{s})/c_0$ on each channel and accumulates a filtered back-projection term,
\begin{align}
\hat{p}_0(\mathbf{r}) \;\propto\; \int_{\mathbf{s}\in\mathcal{S}} w(\mathbf{r},\mathbf{s})
\,\frac{\partial}{\partial t}\big[t\,p(\mathbf{s},t)\big]\bigg|_{t=t^\star}\,dS,
\end{align}
where $p(\mathbf{s},t)$ is the measured pressure and $w(\mathbf{r},\mathbf{s})$ accounts for geometric/sensitivity factors (e.g., $1/R$ spherical spreading and optional obliquity/solid-angle weights for non-closed apertures), following \cite{xu2005universal}. 

\underline{Implementation.} We implemented UBP in C++ with CUDA for GPU acceleration. Each thread processes either (i) a subset of voxels (voxel-driven) or (ii) a subset of detectors (ray-driven); we used a voxel-driven layout for coalesced global memory access to the output volume. Time samples at $t^\star$ are obtained by linear (default) or cubic interpolation of band-limited RF data. To reduce bias in limited-view settings, detector-dependent quadrature weights were precomputed from the local tessellation of $\mathcal{S}$ (Voronoi solid-angle weights). 

\underline{Pre-processing.} Raw time traces underwent (i) DC removal and baseline drift correction, (ii) band-pass filtering matched to the transducer bandwidth, (iii) optional deconvolution of the system impulse response (measured in water) to sharpen the effective temporal point spread, (iv) per-channel gain normalization, and (v) time-zero alignment using the direct-path arrival from a point target (or the water-measurement impulse). The speed of sound $c_0$ was set from the water temperature using a standard polynomial and refined by maximizing image sharpness. 

\underline{Discretization details.} Volumes were reconstructed on a Cartesian grid with voxel pitch chosen to satisfy $\Delta x \lesssim c_0/(2 f_\text{max})$ (Nyquist for the highest usable frequency $f_\text{max}$). We used single-precision accumulation with Kahan compensation to limit summation error; final images were optionally written in 32-bit float. Kernel complexity is $\mathcal{O}(N_v N_d)$ with $N_v$ voxels and $N_d$ detector positions. 

\underline{Output conditioning.} The final $\hat{p}_0$ volumes were apodized to suppress boundary ringing and, where noted, lightly denoised with a divergence-preserving 3-D total variation (TV) post-filter (no edge-sharpening prior to quantitative analyses).

\bmhead{Iterative reconstruction (optimization-based)}
Optimization-based (iterative) PACT reconstructions \cite{xu2002exact,zhang2009effects} were used to compensate for modeling errors, noise, and data incompleteness. We modeled the measured data as
\begin{align}
\mathbf{y} \;=\; \mathbf{H}\,\mathbf{x} \;+\; \boldsymbol{\eta},
\end{align}
where $\mathbf{x}$ is the initial pressure distribution, $\mathbf{H}$ is the forward operator mapping $\mathbf{x}$ to multi-channel time series, and $\boldsymbol{\eta}$ is measurement noise. The forward and adjoint operators were implemented with a time-domain acoustic solver (pseudo-spectral $k$-space or high-order finite differences) with perfectly matched layers (PML) and the measured transducer impulse/receive directivity; the adjoint corresponds to time-reversal wave propagation with the same boundary conditions. 

We solved the composite objective
\begin{align}
\mathbf{x}^\star \;=\; \arg\min_{\mathbf{x}\ge 0}\;
\frac{1}{2}\,\|\mathbf{W}\,(\mathbf{H}\mathbf{x}-\mathbf{y})\|_2^2
\;+\; \lambda\;\mathrm{TV}(\mathbf{x})
\;+\; \frac{\mu}{2}\,\|\mathbf{x}\|_2^2,
\end{align}
where $\mathbf{W}$ whitens the data using the noise power across channels, $\mathrm{TV}$ promotes edge-preserving sparsity, and the Tikhonov term stabilizes high-frequency components in low SNR regimes. We used a first-order primal–dual scheme (Chambolle–Pock) or accelerated proximal gradient (FISTA), with step sizes set from a power-iteration estimate of $\|\mathbf{H}\|_2$ (or from the CFL limit for time-domain solvers). The proximal map for TV used anisotropic shrinkage with optional Huber smoothing for differentiability. Nonnegativity was enforced by projection. 

\underline{Practicalities.} 
(i) \textit{Initialization:} UBP was used as the warm start. 
(ii) \textit{Regularization:} $\lambda$ and $\mu$ were selected by the discrepancy principle (targeting $\mathbb{E}\|\mathbf{W}(\mathbf{H}\mathbf{x}-\mathbf{y})\|_2^2\approx\text{degree of freedom}$) and cross-validated against a sharpness–stability curve; the same $\lambda$ was used across scans unless otherwise stated. 
(iii) \textit{Stopping:} iterations terminated when the relative objective decrease fell below $10^{-3}$ or the gradient norm plateaued; early stopping was preferred in very low SNR. 
(iv) \textit{Subsampling and limited view:} when detectors were non-uniform or sparsely sampled, we incorporated quadrature weights in $\mathbf{H}$ and used TV to mitigate null-space artifacts. 
(v) \textit{Speed-of-sound mismatch:} for mild heterogeneity, we used an effective $c_0$ estimated per scan; for stronger heterogeneity (when applicable), we allowed a spatially varying $c(\mathbf{r})$ in $\mathbf{H}$ while keeping the same adjoint. 

All reconstructions used identical pre-processing, solver tolerances, and boundary settings; only the regularization weights were tuned as noted above. Exact run-time parameters ($\lambda$, optimizer parameters, iteration counts) are tuned for each inference setting.

\bmhead{Reconstruction-then-Denoising network (Denoiser)} We follow the setting of DL-PACT \cite{choi2023DLPACT} for the reconstruction-and-denoising framework. Specifically, the PA wave sensory input $\Psi$ is first fed to UBP for an initial reconstruction, and then fed to a U-Net for denosing, which yields the final reconstruction.
Due to the lack of publicly available code of DL-PACT, we implement a 3D U-Net \cite{ronneberger2015u,cciccek20163d} with residual connections in each convolution block. The method is dubbed as Denoiser. We verify the baseline behaves reasonably by matching the performance reported for the full-sampling and densely-sampling case. Four subsampling and upsampling operations is performed, with a stride of $2\times2\times2$. The channels before the first convolution and subsampling block and after the first, second, third and fourth convolution and subsampling blocks are 32, 64, 128, 256 and 512, respectively. For fairness, the training, validation and test split for the denoising network is the same as the \method. Learning rate, optimizer and weight decay are tuned for the denoising network with the validation set.




\subsection{\method Design}
\label{sec:panodesign}
\bmhead{Overview}

In this section,  calligraphic symbols denote learnable operators acting on function spaces, whereas italic symbols indicate fixed (non-learnable) mappings.  The composite mapping learned by \method\ is
\begin{align}
    \hat{P}=\mathcal{G}_{\bm{\Theta}}(\Psi_k)
        = {\mathcal{U}%
           \bigl(
             \mathcal{F}\!
             \bigl[
               \mathrm{Concat}_{k}\,
               (\mathcal{D}_{k}(\Psi_{k}))
             \bigr]
           \bigr)}
\end{align}
where $\Psi(\theta,\phi,k)\in\mathbb{R}^{N_\theta\times N_\phi\times N_k}$ is the complex-valued frequency–domain pressure measured on a hemispherical detector array \footnote{In practice, real-valued temporal signal is measured and we obtain the complex-valued frequency–domain pressure.} and $\hat{P}\in\mathbb{R}^{N_x\times N_y\times N_z}$ is the reconstructed initial-pressure distribution.  
More rigorously, $\mathcal{G}_{\bm{\Theta}}$ is a neural operator that learns the function space mapping $\psi \mapsto p_0$. In this paper, we considered spatially sampled pressure (measurements) $\Psi=U\psi$ where $U \in H^1(\Omega)\times L^2(\mathbb{R})\mapsto \mathbb{R}^N$ is a sampling operator. We consider different sampling patterns (\cref{fig:teaser}b).

The three ingredients of \method, $\mathcal{U}$ (3D U-Net), $\mathcal{F}$ (FNO), $\mathcal{D}$ (Spherical DISCO),  are detailed below.

\bmhead{Spherical DISCO convolution}
The sensors are spatially distributed over a hemisphere, so an \emph{intrinsically spherical} convolution is essential to preserve neighborhood relationships and guarantee rotational equivariance.  Let $f\colon S^{2}\!\rightarrow\!\mathbb{C}$ be a function on the sphere. 
For a kernel $\kappa$  defined over some compact subset $D \subset \mathbb{R}^d$, the continuous spherical convolution, which transforms input $u$ to output $v$, is given by
\begin{equation}
\label{eq:conv}
    (k \star f)(v) = \int_{D} \kappa(u - v) \cdot f(u) \ \mathrm{d}u
\end{equation}
Given a particular set of input points $(u_j)_{j=1}^m \subset D$ with corresponding quadrature weights $q_j$ and output positions $v_i \in D$, we adopt the discrete-continuous convolutions (DISCO) framework for operator learning~\cite{DISCO,local_no} and approximate the continuous convolution (Eqn. \eqref{eq:conv}) by
\begin{equation}
    \label{eq:disc_sph_conv}
    (k \star g)(v_i) \approx \sum_{j=1}^m \kappa(u_j - v_i) \cdot g(x_j) q_j
\end{equation}
thereby decoupling \emph{resolution-agnostic} learnable parameters from the evaluation grid.  The kernel is parameterised as a finite linear combination of basis functions,
$\kappa=\sum_{\ell=1}^{L}\theta_{\ell}\,\kappa^{\ell}$,
with three complementary bases evaluated in our ablation study: piece-wise linear \cite{local_no}, Haar wavelets and Zernike polynomials (with details on the kernel bases in the following paragraph).  Because Eqn. \eqref{eq:disc_sph_conv} can be executed by resampling the neighbourhood of each detector onto a small equi-angular patch, the operation is accelerated by standard GPU-friendly 2-D convolutions while retaining the equivariance of the underlying continuous operator.

For multi-frequency data, we instantiate $k$ spherical DISCO blocks $\mathcal{D}_{k}$ that share the basis but possess frequency-specific parameters~$\bm{\theta}_{k}$.  Each block outputs a tensor of shape $C\times N_\theta\times N_\phi$, capturing localised spectral–spatial features of $\Psi$.

\underline{Kernel $\kappa$ is defined on a disk for spherical convolutions.}   
A spherical convolution is defined by translating (i.e.\ rotating) a single
template $\kappa$ over the sphere via the left action of $SO(3)$.
To preserve \emph{locality}—a key property of conventional planar
convolutions—we restrict the support of $\kappa$ to the geodesic ball (disk) 
\(
B_r(n)=\{x\in\mathbb S^2\mid d_{\mathbb S^2}(x,n)\le r\},
\)
centred at the north pole $n$. In our experiments, we tune $r$ in the range of $[0.01. 0.15]$ of the input size to balance locality and coverage, and report $r=0.02$ of the input size gives the best validation performance.
Under any rotation $g\in SO(3)$ this ball is carried to another geodesic
ball of identical radius, so the induced operator remains
$SO(3)$-equivariant while never accessing information farther than
$r$ radians away from the evaluation point.

\underline{Kernel bases.}  We visualize the three kernel configurations in \cref{fig:disco}c. 
We specifically consider three different kernel bases: piecewise linear, wavelet, and Zernike.
The bases need to satisfy two properties: (i) Linear Independence: No vector in the set can be expressed as a linear combination of the other vectors within the same set. (ii) Spanning: The set of vectors can be used to represent every other vector in the vector space through linear combinations (scaling and adding them together). 
We use regular wavelet bases \cite{steffen2002theory} and Zernike bases \cite{von1934beugungstheorie} defined on the disk \footnote{Please refer to the Appendix E.2 of \cite{jatyani2025unified} for a detailed illustration of how Zernike bases satisfy the basis properties}. 
Below, we explain how we parameterize the piecewise linear bases.

Denote by $\{(\rho_\ell,\varphi_\ell)\}_{\ell=1}^{L}$ the
collocation points obtained from $K$ concentric rings
$\rho\in\{0,\Delta\rho,\dots,r\}$ and, on each ring
$\rho\!>\!0$, a uniform tessellation in the azimuthal direction
$\varphi\in\{0,\tfrac{2\pi}{M(\rho)},\dots,2\pi\}$.
For every collocation point we attach a separable hat basis
\begin{align}
b_\ell(\rho,\varphi)=
\psi_\text{rad}\!\left(\frac{\rho-\rho_\ell}{\Delta\rho}\right)
\;\psi_\text{ang}\!\left(\frac{\sin\rho_\ell}{\rho_\ell}\,(\varphi-\varphi_\ell)\right),
\end{align}
with one-dimensional linear “tent’’ functions
\(\psi_\text{rad}(t)=\max(1-|t|,0)\) and
\(\psi_\text{ang}(t)=\max(1-|t|,0)\) defined on the periodic interval
$[-\pi,\pi)$.  
The learnable filter is the non-negative linear combination
\(
\kappa(\rho,\varphi)=\sum_{\ell=1}^{L}\theta_\ell\,b_\ell(\rho,\varphi),
\)
whose first four piecewise linear basis functions (for $r=0.1\pi$ and
$L=4$) are shown in the first row of \cref{fig:disco}c.  
This construction yields (i) compact support, (ii) continuous but
anisotropic angular response, and (iii) a sparse
evaluation matrix $K_{ij}=\kappa\!\big(d_{\mathbb S^2}(g_i^{-1}x_j)\big)$
amenable to efficient DISCO implementation.



\bmhead{Fourier neural operator (FNO) for global feature learning}
Local spherical convolutions provide only a limited receptive field.  
To propagate information across the entire detector dome we employ a Fourier Neural Operator (FNO) \cite{li2020fourier} that acts \emph{spectrally} on the angular coordinates while treating frequency channels as an additional depth dimension. FNO is chosen because it is a powerful neural operator
framework that efficiently learns mappings in function spaces, with many applications as surrogate models
for solving partial differential equations (PDEs) for computational imaging tasks \cite{wang2025ultrasound,tolooshams2025vars,jatyani2025unified,tolooshams2025vars}. 

\underline{3D FNO.}
Let  
\(
f^{(0)}
   =\mathrm{Concat}_{k}\!\bigl(\mathcal{D}_{k}(\Psi_k)\bigr)
   \in\mathbb{C}^{C\times N_\theta\times N_\phi\times N_k},
\)  
where $C$ is the number of feature channels produced by the DISCO encoder,  
$(N_\theta,N_\phi)$ denote the angular grid and $N_k$ the number of
wavenumbers.  We also denote the Fourier transform as $F$.
The FNO refines $f^{(0)}$ through $L$ spectral layers,
each of which performs four steps:

\begin{enumerate}[label=(\roman*),leftmargin=4em]
\item {Spatial FFT:}
      \(\hat{f}^{(\ell-1)}={F}_{\theta,\phi}\bigl[f^{(\ell-1)}\bigr]\)
      is computed for every $(c,k)$ slice, leaving the
      wavenumber axis $k$ unchanged. ${F}_{\theta,\phi}$ denotes the 2D Fourier transform on $(\theta,\phi)$ dimension, as the last dimension is already in the frequency space of the time.
\item {Spectral convolution:}
      The complex spectrum is modulated by a
      learnable tensor  
      \(M^{(\ell)}\in\mathbb{C}^{C\times J_\theta\times J_\phi\times J_k}\)  
      restricted to the lowest
      $\lvert\xi_\theta\rvert\le J_\theta$,
      $\lvert\xi_\phi\rvert\le J_\phi$, $\lvert\xi_k\rvert\le J_k$ modes:
\begin{align}
      \tilde{f}^{(\ell)}
        =\sum_{\lvert\xi_\theta\rvert\le J_\theta, \lvert\xi_\phi\rvert\le J_\phi, \lvert\xi_k\rvert\le J_k}
          M^{(\ell)}_{\xi_\theta,\xi_\phi,\xi_k}\;
          \hat{f}^{(\ell-1)}_{\xi_\theta,\xi_\phi,\xi_k}.
\end{align}
\item {Inverse FFT:}
      \(g^{(\ell)}
         ={F}^{-1}_{\theta,\phi}
            \bigl[\tilde{f}^{(\ell)}\bigr]\).
\item {Point-wise non-linearity:}
      \(f^{(\ell)}
         =\sigma\!\bigl(B^{(\ell)}g^{(\ell)} + b^{(\ell)}\bigr)\),
      where \(B^{(\ell)}\) is a $1\times1\times1$  convolution 
      shared across $(\theta,\phi, k)$ and  
      \(\sigma\) is the ReLU activation.
\end{enumerate}

We set $(J_\theta,J_\phi, J_k)=(13, 22, 98)$ in our experiment, which sufficed to capture global context without incurring prohibitive memory cost. The number of cutoff modes is tuned by the characteristics of the input data and the available computational resources.

\underline{Convert back to temporal signal.}
The output of the final layer,
\(f^{(L)}\in\mathbb{C}^{C\times N_\theta\times N_\phi\times N_k}\),
is converted back to the time domain by an inverse FFT along $k$,
\begin{align}
z={F}^{-1}_{k}\!\bigl[f^{(L)}\bigr]\in
  \mathbb{R}^{C\times N_\theta\times N_\phi\times N_t},
\end{align}
after which the negligible imaginary residue is discarded.  
The real tensor $z$ serves as the input to the subsequent 3-D U-Net decoder, providing a globally coherent yet high-resolution estimate of the photo-acoustic field.

\bmhead{Multi-scale refinement with 3D \mbox{U-Net}}
3D FNO learns in the low-frequency space and may not reconstruct high-frequency components of the image.
We therefore refine the intermediate prediction with a lightweight residual 3D U-Net \cite{ronneberger2015u,cciccek20163d}.  Three down- and up-sampling stages with kernel size $3\times 3\times 3$ and stride $2\times 2 \times 2$ progressively aggregate contextual information and then reinject it via skip connections, resulting in sharper edges and improved tissue contrast.  Encoder channel widths of $(16, 32, 64)$ balance accuracy and memory footprint.

\bmhead{Physics-aware learning}

Purely data-driven supervision is prone to “hallucinating’’ anatomically plausible but acoustically
infeasible structures.  
To constrain \method\ we therefore augment the voxel-wise loss by an \emph{explicit} enforcement of the governing wave equation,
\begin{equation}
\label{eq:loss_total}
\mathcal{L}(\bm{\Theta})=
\lambda_{\mathrm{img}}\,
        \bigl\lVert \hat{P}-P\bigr\rVert_{1}
\;+\;
\lambda_{\mathrm{phys}}\,
        \bigl\lVert MA\hat{P}-M\Psi\bigr\rVert_{2}^{2},
\end{equation}
where $A:\mathbb{R}^{N_x\times N_y\times N_z}\!\rightarrow\!\mathbb{R}^{N_\theta\times N_\phi\times N_k}$ is the forward
photo-acoustic operator. $M$ is a random mask at different training iterations that makes the training faster and increase the robustness with the randomness. Empirically, for each iteration, we randomly sample 15 modes and 40 sensors in the sensor array to check the validity of the physics -- this number is tuned to balance the trade-off between GPU memory limits, training speed and model performance. Hyperparameter $\lambda_{\mathrm{img}}, \lambda_{\mathrm{phys}}$ is tuned in the validation set. 
We find $\lambda_{\mathrm{img}}=1, \lambda_{\mathrm{phys}}=0.5$ give the best validation performance.


\textbf{}

\subsection{Implementation Details}
\label{sec:implementation}
\bmhead{Data}
The resulting dataset indices and splits are summarized in Table~\ref{tab:data_comp}; reconstruction uses identical physics/filters as detailed in the Methods.

\begin{table}[h]
\centering
\footnotesize
\caption{Data used in the study. Note that this refers to the unique number of 3D images, not considering the data augmentation during training and fine-tuning.}
    \setlength{\tabcolsep}{2pt}
  \begin{tabular}{llllllll}
 \toprule

                                & \multicolumn{4}{c}{Simulation Set}                                            & \multicolumn{3}{c}{Ex-vivo Set}                 \\  \cmidrule{2-8} 
                                & Training                 & Validation               & Evaluation       &Sum        & Fine-Tuning            & Evaluation     &Sum         \\ \midrule
Number of data samples          & 7,000	& 1,000	&	2,000	&	10,000	&	37	&	14		&51 \\
\bottomrule
\end{tabular}

          \label{tab:data_comp}
\end{table}

\bmhead{Training details} The model is implemented with the PyTorch framework. We first train \method on simulated data for 90 epochs using the Adam optimizer~\cite{kingma2014adam} with learning rate $\eta = 0.002$ and $\beta = (0.9, 0.999)$. The effective batch size is 40 (achieved via gradient accumulation every 10 iterations). We then fine-tune the model on real data for another 10 epochs using Adam with a reduced learning rate of $0.1\eta = 0.0002$ and $\beta = (0.9, 0.999)$. In total, training took 1.5 days on one NVIDIA A100 GPU.
\bmhead{Data augmentation}
To improve reconstruction performance and ensure fair comparison across learning-based methods, we apply the same two augmentation strategies to \method and all learning-based baselines during training.
First, random white Gaussian noise of 5--20\,dB is added to the RF measurement signal $\Psi$ on-the-fly at each training iteration, improving robustness to sensor noise.
Second, with probability 0.2 per iteration, we combine two training samples by summing both their RF signals and their corresponding 3D images: $\Psi' = \Psi_1 + \Psi_2$, $P' = P_1 + P_2$.  This mixing is physically valid because the PACT forward operator $A$ is linear ($A(P_1 + P_2) = \Psi_1 + \Psi_2$), so the combined pair is a legitimate training example.  The strategy increases dataset variability and encourages the model to handle superimposed structures.

\bmhead{Sim-to-Real domain adaptation}
To make \method work for real data, we apply the following transformation on the simulation data to reduce the simulation and real data's domain gap.
1) Applying sensor-specific amplification rescaling. This is applied because the real PACT system has a sensor-specific amplification scale due to manufacturing tolerances; we calibrate and measure these scales and apply them to the simulated data to reduce the domain gap. Note that noise augmentation (5--20\,dB, described above) is applied during all training, including the fine-tuning stage, and serves dual purposes: general robustness and sim-to-real transfer.
2) Finetuning on real data. As mentioned earlier, after training on the simulation data, \method is also fine-tuned on a small amount of real data to improve the real performance. Considering the size of the real data and to avoid forgetting, we use a mixed data training strategy (75\% simulated data + 25\% real data) for each training iteration. 

\bmhead{Evaluation Protocols}

We adopt several metrics to evaluate the 3D image reconstruction performance of \method.
\begin{enumerate}
\item Cosine similarity: The 3D volumetric images are first normalized with $\ell_2$ norm of 1, and then calculated for the cosine similarity, i.e.  $\text{cosine\_similarity}(P, \hat{P}) = \frac{P\cdot \hat{P}}{\|P\|_2 \|\hat{P}\|_2}$.Cosine similarity normalizes the image before comparing the similarity, thus eliminating the effect of different scales of the PA magnitude. The normalization is necessary as the relative magnitude is meaningful in PACT reconstruction. 
\item The Peak Signal-to-Noise Ratio (PSNR) measures the ratio between the maximum possible power of a signal and the power of corrupting noise that affects the fidelity of its representation:
$\text{PSNR} = 10 \cdot \log_{10} \left( \frac{\max(P_{i,j,k})^2}{\frac{1}{HWD} \sum_{i,j,k} \left( P_{i,j,k}  - \hat{P}_{i,j,k}  \right)^2} \right)$.
Here $H$, $W$, and $D$ are the height, width, and depth of the volume, respectively, and $P_{i,j,k}$, $\hat{P}_{i,j,k}$ denote the $(i,j,k)$-th voxel of the ground-truth and reconstructed volumes.
    \item The Normalized Mean Squared Error (NMSE) measures the mean squared error normalized by the ground-truth volume's energy:
$\text{NMSE} = \frac{\displaystyle\sum_{i,j,k} \left( P_{i,j,k} - \hat{P}_{i,j,k} \right)^2}{\displaystyle\sum_{i,j,k} \left( P_{i,j,k} \right)^2}$.
\end{enumerate}

\clearpage

\subsection*{Code Availability}

The code is publicly available at 
$\texttt{https://github.com/neuraloperator/pact3d}$.

\subsection*{Data Availability}
We provide the source code for generating the synthetic dataset as in\\ 
$\texttt{https://github.com/neuraloperator/pact3d/tree/main/data\_gen}$.
Due to the large size of the PA wave data, the datasets generated and analysed during the current study are available from the corresponding author on reasonable request.
\subsection*{Acknowledgements}

This work is supported in part by ONR (MURI grants N000142312654 and N000142012786) and the United States National Institutes of Health (NIH) grants U01 EB029823 (BRAIN Initiative), R35 CA220436 (Outstanding Investigator Award), and R01 CA282505.
J.W. is supported in part by the Pritzker AI+Science initiative and Schmidt Sciences. A.A. is supported in part by the Bren endowed chair and the AI2050 senior fellow program at Schmidt Sciences. Z. Li is supported in part by the NVIDIA Fellowship. The authors thank Rui Cao for helpful discussions.

\subsection*{Author Contributions}
All authors have contributed to the publication.
J.W. implemented the main codebase for the deep learning part, with the help of Y.A and C.W. Y.A. and J.W. implemented the main codebase for the numerical simulation. Y.A. and A.K. implemented the baseline methods for the inverse problem, with the help of J.W. The computations were undertaken by J.W., with the help of C.W., B.B. and Z.L.
The phantom experiments were undertaken by Y.Z., K.S. and Y.L. The phantom experimental data preprocessing was undertaken by Y.Z. and Y.A. 
The main text was initially drafted by J.W. with contributions, editing and comments from all authors, particularly Y.A. and K.S. The Supplementary was primarily written by J.W. 
J.W. created all main text figures and tables in consultation with K.A., L.V.W. and A.A.
 The scientific findings were discussed and approved by all contributors.
 
\subsection*{Competing Interests}
L.V.W. has a financial interest in Microphotoacoustics Inc., CalPACT LLC, and Union Photoacoustic Technologies Ltd., which, however, did not support this work. The rest of the authors declare that they have no competing interests.

\subsection*{Additional Information}

Supplementary Information is available for this paper.

Correspondence and requests for materials should be addressed to Lihong V. Wang and Anima Anandkumar.

\bibliography{references.bib}

\newpage

\section*{Supplementary Information}

\begin{appendices}

\section{Numerical Results}
We present numerical results of the proposed and existing methods as a supplement to the main figures.
Unless otherwise noted, all metrics are computed against fully sampled ground-truth volumes. Cosine similarity (\%) and PSNR (dB) are higher-is-better, while NMSE is lower-is-better. An acceleration of $r{\times}$ indicates that only $1/r$ of the sensors are used/uniformly sampled relative to the fully sampled acquisition. All comparisons are conducted under identical preprocessing, evaluation protocols, and test splits to ensure fairness.

\subsection{Comparison with Baselines}

\bmhead{Performance under different subsampling rates}
The subsampling pattern is kept as uniform for controlled experiments. Both simulated and real results are presented in \cref{fig:sim_results}b and \cref{fig:real_results}b of the main paper. Below, we present the complete numerical results on the simulated data (Table~\ref{tab:sim_results}) and real phantom data (Table~\ref{tab:real_results}). Note that the baseline Denoiser \cite{choi2023DLPACT} slightly outperforms \method at $6{\times}$ acceleration on simulated data (90.3\% vs. 89.9\%), but \method outperforms the Denoiser at all other subsampling rates and on real phantom data at all rates, indicating Denoiser may suffer stronger overfitting to a specific sampling rate. The performance gap widens with increasing acceleration, confirming the robustness of direct inverse operator learning to sparse measurements.

\underline{Simulated data.} On simulated data, \method matches Denoiser at $6{\times}$ (--0.4 percentage points cosine similarity) and surpasses it at all higher subsampling rates: $+2.7$, $+5.5$, and $+14.4$ percentage points at $10{\times}$/$15{\times}$/$20{\times}$, respectively. PSNR increases by $+0.7$, $+1.4$, and $+1.8$\,dB at $10{\times}$/$15{\times}$/$20{\times}$, and NMSE is reduced by $15.4\%$, $26.8\%$, and $34.0\%$ (tie at $6{\times}$). Notably, the quality of \method degrades more gracefully with increasing acceleration: cosine similarity drops only $12.8$ percentage points from $6{\times}$ to $20{\times}$ for \method, versus $27.6$ percentage points for the Denoiser. This confirms that direct inverse operator learning is substantially more robust to sparse measurements than the two-step reconstruct-then-denoise paradigm.

\underline{Real phantom data.} On real phantoms, \method improves over the Denoiser by $+4.3$, $+3.0$, $+28.4$, and $+9.6$ percentage points in cosine similarity at $6{\times}$/$10{\times}$/$15{\times}$/$20{\times}$, respectively, with corresponding PSNR gains of $+1.8$, $+1.5$, $+3.2$, and $+1.6$\,dB. NMSE decreases by $33.3\%$, $25.0\%$, $53.8\%$, and $31.3\%$ at the four rates. The largest gap at $15{\times}$ highlights the robustness of \method to severe sparsity in real acquisition settings. These trends are consistent with the simulated results, validating the sim-to-real generalizability of \method.

\begin{table}[htb]
\centering
\caption{All methods' performance on simulated data under uniform subsampling at varying subsampling rates. Cosine similarity (\%) and PSNR (dB) are higher-is-better; NMSE is lower-is-better. \method outperforms UBP at all subsampling rates. Notably, while \method and the Denoiser perform comparably at $6{\times}$, \method progressively outperforms the Denoiser at higher acceleration, with the largest margin at $20{\times}$.}
\begin{tabular}{l|llll|llll|llll}
 \toprule
\multirow{2}{*}{\textbf{method}} & \multicolumn{4}{c|}{\textbf{Cosine similarity (\%)}} & \multicolumn{4}{c|}{\textbf{PSNR}} & \multicolumn{4}{c}{\textbf{NMSE}} \\
                                 & 6x          & 10x         & 15x        & 20x        & 6x     & 10x    & 15x    & 20x    & 6x     & 10x    & 15x    & 20x    \\ \midrule
\textbf{UBP}~\cite{xu2005universal}                     & 65.1        & 53.9        & 46.0       & 39.7       & 18.3   & 17.2   & 16.6   & 16.2   & 0.0148 & 0.0190 & 0.0218 & 0.0240 \\
\textbf{Denoiser}~\cite{choi2023DLPACT}                    & 90.3        & 85.4        & 76.2       & 62.7       & 25.0   & 24.1   & 22.5   & 20.3   & 0.0032 & 0.0039 & 0.0056 & 0.0094 \\
{\bf \method}                      & 89.9        & 88.1        & 81.7       & 77.1       & 24.9   & 24.8   & 23.9   & 22.1   & 0.0032 & 0.0033 & 0.0041 & 0.0062 \\
\bottomrule
\end{tabular}
          \label{tab:sim_results}
\end{table}

\begin{table}[htb]
\centering
\caption{All methods' performance on real phantom data under uniform subsampling at varying subsampling rates. Metrics follow the same conventions as Table~\ref{tab:sim_results}. The gain of \method over competing methods is especially pronounced at $15{\times}$ acceleration, where the Denoiser's performance collapses while \method retains high fidelity.}
\begin{tabular}{l|llll|llll|llll}
 \toprule
\multirow{2}{*}{\textbf{method}} & \multicolumn{4}{c|}{\textbf{Cosine similarity (\%)}} & \multicolumn{4}{c|}{\textbf{PSNR}} & \multicolumn{4}{c}{\textbf{NMSE}} \\
                                 & 6x          & 10x         & 15x        & 20x        & 6x     & 10x    & 15x    & 20x    & 6x     & 10x    & 15x    & 20x    \\ \midrule
\textbf{UBP}~\cite{xu2005universal}                     & 77.3        & 75.1         & 42.1         & 31.0         & 30.7        & 30.4         & 28.8         & 27.9         & 0.0008      & 0.0009       & 0.0013       & 0.0016       \\
\textbf{Denoiser}~\cite{choi2023DLPACT}                    & 83.0        & 78.6         & 44.5         & 31.6         & 35.5        & 34.1         & 28.9         & 27.9         & 0.0003      & 0.0004       & 0.0013       & 0.0016       \\
{\bf \method}                      & 87.3        & 81.6         & 72.9         & 41.2         & 37.3        & 35.6         & 32.1         & 29.5         & 0.0002      & 0.0003       & 0.0006       & 0.0011      \\
\bottomrule
\end{tabular}
          \label{tab:real_results}
\end{table}




\bmhead{Performance under different subsampling patterns}
To evaluate the generality of \method beyond uniform acceleration, we compare all methods under three distinct subsampling patterns at a fixed 3$\times$ subsampling rate: uniform subsampling, limited-angle acquisition in azimuth, and limited-angle acquisition in elevation (see Fig.~1b in the main paper for pattern illustrations). The numerical results are presented in Table~\ref{tab:dpattern}. \method achieves the best cosine similarity in all three patterns. The advantage is most pronounced for the limited-angle patterns, where both UBP and the iterative solver suffer from substantial artifacts due to the restricted measurement aperture. The Denoiser improves over the solvers but cannot fully compensate for incomplete angular coverage. In contrast, \method, by learning the direct inverse operator with physics regularization, reconstructs coherent 3D structures even from highly constrained acquisition geometries.
\begin{table}[htb]
\centering
\caption{Performance of all methods under different subsampling patterns at a fixed $3{\times}$ subsampling rate. Metric is cosine similarity (\%), higher is better. \method consistently achieves the best performance across all three acquisition patterns. The limited-angle settings expose the largest performance gap between end-to-end operator learning and two-step methods.}
  \begin{tabular}{llll}
 \toprule

     {\bf subsampling pattern at 3$\times$ }     & {\bf Uniform} & {\bf Limited angle (azimuth)} & {\bf Limited angle (elevation)}  \\ \midrule
{\bf UBP}~\cite{xu2005universal} &67.1	&49.4	&46.4\\
{\bf Iterative solver}~\cite{xu2002exact} &78.5	&56.9	&55.2\\
{\bf Denoiser}~\cite{choi2023DLPACT} &95.0	&79.4	&75.0\\
{\bf \method} & 92.6	&86.4	&86.0\\
\bottomrule
\end{tabular}
          \label{tab:dpattern}
\end{table}

\subsection{Ablation Study on the Physics Loss}
Table~\ref{tab:physics_loss} reports the effect of removing the physics consistency loss during training. For $6{\times}$, $10{\times}$, $15{\times}$, and $20{\times}$ uniform subsampling, adding the physics loss yields cosine similarity gains of 2.7\%, 3.2\%, 4.6\%, and 5.0\%, respectively---an average improvement of 3.9 percentage points. As a reference, the corresponding training convergence curves are shown in \cref{fig:ablation}e of the main paper.

The benefit of the physics loss grows with the acceleration factor. Intuitively, as fewer measurements are available, the inverse problem becomes more ill-posed and data-only supervision can no longer uniquely constrain the reconstruction. The physics loss regularizes predictions toward Helmholtz-equation-consistent fields, preserving high-frequency structures that would otherwise be lost. This makes the physics loss especially important in the clinically relevant high-acceleration regime (${\geq}15{\times}$).

Across all tested settings, \method with physics loss maintains the best trade-off between reconstruction fidelity and measurement sparsity, reinforcing the value of embedding forward-physics constraints directly into the learning objective.

 
\begin{table}[htb]
\centering
\caption{Ablation study on the physics loss under uniform subsampling. The metric is cosine similarity (\%), higher is better. Adding the physics loss yields an average gain of 3.9 percentage points across all tested subsampling rates, with the benefit increasing at higher acceleration where the inverse problem becomes more ill-posed.}
  \begin{tabular}{lllll}
 \toprule

     subsampling rate (uniform)         & 6x   & 10x  & 15x  & 20x  \\ \midrule
              \method w/o physics loss & 87.2 & 84.8 & 77.1 & 72.1\\
\method          & 89.9\textcolor{green}{($\uparrow$2.7)}   & 88.1\textcolor{green}{($\uparrow$3.2)} & 81.7\textcolor{green}{($\uparrow$4.6)} & 77.1\textcolor{green}{($\uparrow$5.0)} \\
\bottomrule
\end{tabular}

          \label{tab:physics_loss}
\end{table}

\end{appendices}

\end{document}